\documentclass[longauth]{aa} 
\usepackage[varg]{txfonts}
\usepackage{natbib}
\usepackage[colorlinks=true, allcolors=blue]{hyperref}
\usepackage{orcidlink}

\usepackage{upgreek}
\usepackage{graphicx}   

\begin{document}

\title{Studying geometry of the ultraluminous X-ray pulsar Swift~J0243.6+6124 using X-ray and optical polarimetry}
\titlerunning{A\&A, 691, A123 (2024)}
\authorrunning{Poutanen, J., et al.}

\author{\small  Juri~Poutanen \inst{\ref{in:UTU}}\orcidlink{0000-0002-0983-0049}
\and Sergey~S.~Tsygankov \inst{\ref{in:UTU}}\orcidlink{0000-0002-9679-0793}
\and Victor~Doroshenko\inst{\ref{in:Tub}}\orcidlink{0000-0001-8162-1105} 
\and Sofia~V.~Forsblom \inst{\ref{in:UTU}}\orcidlink{0000-0001-9167-2790}
\and Peter~Jenke \inst{\ref{in:UAH}}\orcidlink{0000-0002-9877-6768} 
\and Philip~Kaaret \inst{\ref{in:NASA-MSFC}}\orcidlink{0000-0002-3638-0637} 
\and Andrei~V.~Berdyugin \inst{\ref{in:UTU}}\orcidlink{0000-0002-9353-5164} 
\and Dmitry~Blinov
\inst{\ref{in:AstroCrete},\ref{in:PhysCrete}}\orcidlink{0000-0003-0611-5784}
\and Vadim~Kravtsov \inst{\ref{in:UTU}}\orcidlink{0000-0002-7502-3173} 
\and Ioannis~Liodakis \inst{\ref{in:NASA-MSFC},\ref{in:AstroCrete}}\orcidlink{0000-0001-9200-4006}
\and Anastasia~Tzouvanou \inst{\ref{in:PhysCrete}}\orcidlink{0009-0004-4171-8454} 
\and Alessandro~Di~Marco \inst{\ref{in:INAF-IAPS}}\orcidlink{0000-0003-0331-3259} 
\and Jeremy~Heyl \inst{\ref{in:UBC}}\orcidlink{0000-0001-9739-367X} 
\and Fabio~La~Monaca \inst{\ref{in:INAF-IAPS},\ref{in:UniRoma2},\ref{in:LaSapienza}}\orcidlink{0000-0001-8916-4156}  
\and Alexander~A.~Mushtukov  \inst{\ref{in:Oxford}}\orcidlink{0000-0003-2306-419X}  
\and George~G.~Pavlov \inst{\ref{in:PSU}}\orcidlink{0000-0002-7481-5259}   
\and Alexander~Salganik \inst{\ref{in:SPBU},\ref{in:IKI}}\orcidlink{0000-0003-2609-8838}   
\and Alexandra~Veledina \inst{\ref{in:UTU},\ref{in:Nordita}}\orcidlink{0000-0002-5767-7253}
\and Martin~C.~Weisskopf \inst{\ref{in:NASA-MSFC}}\orcidlink{0000-0002-5270-4240}   
\and Silvia~Zane  \inst{\ref{in:MSSL}}\orcidlink{0000-0001-5326-880X}  
\and Vladislav~Loktev \inst{\ref{in:UTU}}\orcidlink{0000-0001-6894-871X}  
\and Valery~F.~Suleimanov\inst{\ref{in:Tub}}\orcidlink{0000-0003-3733-7267}  
\and Colleen~Wilson-Hodge  \inst{\ref{in:NASA-MSFC}}\orcidlink{0000-0002-8585-0084}   
\and Svetlana~V.~Berdyugina\inst{\ref{in:IRSOL},\ref{in:KIS}}\orcidlink{0000-0002-2238-7416}  
\and Masato~Kagitani \inst{\ref{in:Tohoku}}\orcidlink{0000-0002-4115-7122}    
\and Vilppu~Piirola \inst{\ref{in:UTU}}\orcidlink{0000-0003-0186-206X}     
\and Takeshi~Sakanoi \inst{\ref{in:Tohoku}}\orcidlink{0000-0002-7146-9020}   
\and Iv\'an~Agudo \inst{\ref{in:CSIC-IAA}}\orcidlink{0000-0002-3777-6182}  
\and Lucio~A.~Antonelli \inst{\ref{in:INAF-OAR},\ref{in:ASI-SSDC}}\orcidlink{0000-0002-5037-9034}   
\and Matteo~Bachetti \inst{\ref{in:INAF-OAC}}\orcidlink{0000-0002-4576-9337}   
\and Luca~Baldini  \inst{\ref{in:INFN-PI},      \ref{in:UniPI}}\orcidlink{0000-0002-9785-7726}   
\and Wayne~H.~Baumgartner  \inst{\ref{in:NASA-MSFC}}\orcidlink{0000-0002-5106-0463}  
\and Ronaldo~Bellazzini  \inst{\ref{in:INFN-PI}}\orcidlink{0000-0002-2469-7063}   
\and Stefano~Bianchi \inst{\ref{in:UniRoma3}}\orcidlink{0000-0002-4622-4240}  
\and Stephen~D.~Bongiorno \inst{\ref{in:NASA-MSFC}}\orcidlink{0000-0002-0901-2097}   
\and Raffaella~Bonino  \inst{\ref{in:INFN-TO},\ref{in:UniTO}}\orcidlink{0000-0002-4264-1215}  
\and Alessandro~Brez  \inst{\ref{in:INFN-PI}}\orcidlink{0000-0002-9460-1821}   
\and Niccol\`{o}~Bucciantini 
\inst{\ref{in:INAF-Arcetri},\ref{in:UniFI},\ref{in:INFN-FI}}\orcidlink{0000-0002-8848-1392}  
\and Fiamma~Capitanio \inst{\ref{in:INAF-IAPS}}\orcidlink{0000-0002-6384-3027} 
\and Simone~Castellano \inst{\ref{in:INFN-PI}}\orcidlink{0000-0003-1111-4292}   
\and Elisabetta~Cavazzuti \inst{\ref{in:ASI}}\orcidlink{0000-0001-7150-9638}   
\and Chien-Ting~Chen \inst{\ref{in:USRA-MSFC}}\orcidlink{0000-0002-4945-5079}  
\and Stefano~Ciprini \inst{\ref{in:INFN-Roma2},\ref{in:ASI-SSDC}}\orcidlink{0000-0002-0712-2479}  
\and Enrico~Costa \inst{\ref{in:INAF-IAPS}}\orcidlink{0000-0003-4925-8523}   
\and Alessandra~De~Rosa \inst{\ref{in:INAF-IAPS}}\orcidlink{0000-0001-5668-6863}  
\and Ettore~Del~Monte \inst{\ref{in:INAF-IAPS}}\orcidlink{0000-0002-3013-6334}   
\and Laura~Di~Gesu \inst{\ref{in:ASI}}\orcidlink{0000-0002-5614-5028}   
\and Niccol\`{o}~Di~Lalla \inst{\ref{in:Stanford}}\orcidlink{0000-0002-7574-1298}  
\and Immacolata~Donnarumma \inst{\ref{in:ASI}}\orcidlink{0000-0002-4700-4549}  
\and Michal~Dov\v{c}iak \inst{\ref{in:CAS-ASU}}\orcidlink{0000-0003-0079-1239}
\and Steven~R.~Ehlert \inst{\ref{in:NASA-MSFC}}\orcidlink{0000-0003-4420-2838}    
\and Teruaki~Enoto \inst{\ref{in:RIKEN}}\orcidlink{0000-0003-1244-3100}  
\and Yuri~Evangelista \inst{\ref{in:INAF-IAPS}}\orcidlink{0000-0001-6096-6710}  
\and Sergio~Fabiani \inst{\ref{in:INAF-IAPS}}\orcidlink{0000-0003-1533-0283}  
\and Riccardo~Ferrazzoli \inst{\ref{in:INAF-IAPS}}\orcidlink{0000-0003-1074-8605}   
\and Javier~A.~Garcia \inst{\ref{in:GSFC}}\orcidlink{0000-0003-3828-2448}
\and Shuichi~Gunji \inst{\ref{in:Yamagata}}\orcidlink{0000-0002-5881-2445}   
\and Kiyoshi~Hayashida \inst{\ref{in:Osaka}}\thanks{Deceased.}  
\and Wataru~Iwakiri \inst{\ref{in:Chiba}}\orcidlink{0000-0002-0207-9010}   
\and Svetlana~G.~Jorstad \inst{\ref{in:BU},\ref{in:SPBU}}\orcidlink{0000-0001-9522-5453}   
\and Vladimir~Karas \inst{\ref{in:CAS-ASU}}\orcidlink{0000-0002-5760-0459}  
\and Fabian~Kislat \inst{\ref{in:UNH}}\orcidlink{0000-0001-7477-0380}   
\and Takao~Kitaguchi  \inst{\ref{in:RIKEN}} 
\and Jeffery~J.~Kolodziejczak \inst{\ref{in:NASA-MSFC}}\orcidlink{0000-0002-0110-6136}   
\and Luca~Latronico  \inst{\ref{in:INFN-TO}}\orcidlink{0000-0002-0984-1856}   
\and Simone~Maldera \inst{\ref{in:INFN-TO}}\orcidlink{0000-0002-0698-4421}    
\and Alberto~Manfreda \inst{\ref{INFN-NA}}\orcidlink{0000-0002-0998-4953}  
\and Fr\'{e}d\'{e}ric~Marin \inst{\ref{in:Strasbourg}}\orcidlink{0000-0003-4952-0835}   
\and Andrea~Marinucci \inst{\ref{in:ASI}}\orcidlink{0000-0002-2055-4946}   
\and Alan~P.~Marscher \inst{\ref{in:BU}}\orcidlink{0000-0001-7396-3332}   
\and Herman~L.~Marshall \inst{\ref{in:MIT}}\orcidlink{0000-0002-6492-1293}  
\and Francesco~Massaro \inst{\ref{in:INFN-TO},\ref{in:UniTO}}\orcidlink{0000-0002-1704-9850}   
\and Giorgio~Matt  \inst{\ref{in:UniRoma3}}\orcidlink{0000-0002-2152-0916}  
\and Ikuyuki~Mitsuishi \inst{\ref{in:Nagoya}}
\and Tsunefumi~Mizuno \inst{\ref{in:Hiroshima}}\orcidlink{0000-0001-7263-0296}   
\and Fabio~Muleri \inst{\ref{in:INAF-IAPS}}\orcidlink{0000-0003-3331-3794}
\and Michela~Negro \inst{\ref{in:LSU}}\orcidlink{0000-0002-6548-5622} 
\and Chi-Yung~Ng \inst{\ref{in:HKU}}\orcidlink{0000-0002-5847-2612}  
\and Stephen~L.~O'Dell \inst{\ref{in:NASA-MSFC}}\orcidlink{0000-0002-1868-8056}    
\and Nicola~Omodei \inst{\ref{in:Stanford}}\orcidlink{0000-0002-5448-7577}  
\and Chiara~Oppedisano \inst{\ref{in:INFN-TO}}\orcidlink{0000-0001-6194-4601}    
\and Alessandro~Papitto \inst{\ref{in:INAF-OAR}}\orcidlink{0000-0001-6289-7413}  
\and Abel~L.~Peirson \inst{\ref{in:Stanford}}\orcidlink{0000-0001-6292-1911}  
\and Matteo~Perri \inst{\ref{in:ASI-SSDC},\ref{in:INAF-OAR}}\orcidlink{0000-0003-3613-4409}  
\and Melissa~Pesce-Rollins \inst{\ref{in:INFN-PI}}\orcidlink{0000-0003-1790-8018}   
\and Pierre-Olivier~Petrucci \inst{\ref{in:Grenoble}}\orcidlink{0000-0001-6061-3480}
\and Maura~Pilia \inst{\ref{in:INAF-OAC}}\orcidlink{0000-0001-7397-8091}   
\and Andrea~Possenti \inst{\ref{in:INAF-OAC}}\orcidlink{0000-0001-5902-3731}   
\and Simonetta~Puccetti \inst{\ref{in:ASI-SSDC}}\orcidlink{0000-0002-2734-7835}  
\and Brian~D.~Ramsey \inst{\ref{in:NASA-MSFC}}\orcidlink{0000-0003-1548-1524}    
\and John~Rankin \inst{\ref{in:INAF-IAPS}}\orcidlink{0000-0002-9774-0560}   
\and Ajay~Ratheesh \inst{\ref{in:INAF-IAPS}}\orcidlink{0000-0003-0411-4243} 
\and Oliver~J.~Roberts \inst{\ref{in:USRA-MSFC}}\orcidlink{0000-0002-7150-9061}  
\and Roger~W.~Romani \inst{\ref{in:Stanford}}\orcidlink{0000-0001-6711-3286}  
\and Carmelo~Sgr\`o \inst{\ref{in:INFN-PI}}\orcidlink{0000-0001-5676-6214}    
\and Patrick~Slane \inst{\ref{in:CfA}}\orcidlink{0000-0002-6986-6756}    
\and Paolo~Soffitta \inst{\ref{in:INAF-IAPS}}\orcidlink{0000-0002-7781-4104}   
\and Gloria~Spandre \inst{\ref{in:INFN-PI}}\orcidlink{0000-0003-0802-3453}   
\and Douglas~A.~Swartz \inst{\ref{in:USRA-MSFC}}\orcidlink{0000-0002-2954-4461}  
\and Toru~Tamagawa \inst{\ref{in:RIKEN}}\orcidlink{0000-0002-8801-6263}  
\and Fabrizio~Tavecchio \inst{\ref{in:INAF-OAB}}\orcidlink{0000-0003-0256-0995}  
\and Roberto~Taverna \inst{\ref{in:UniPD}}\orcidlink{0000-0002-1768-618X}  
\and Yuzuru~Tawara \inst{\ref{in:Nagoya}} 
\and Allyn~F.~Tennant \inst{\ref{in:NASA-MSFC}}\orcidlink{0000-0002-9443-6774}    
\and Nicholas~E.~Thomas \inst{\ref{in:NASA-MSFC}}\orcidlink{0000-0003-0411-4606}    
\and Francesco~Tombesi  \inst{\ref{in:UniRoma2},\ref{in:INFN-Roma2}}\orcidlink{0000-0002-6562-8654}  
\and Alessio~Trois \inst{\ref{in:INAF-OAC}}\orcidlink{0000-0002-3180-6002}  
\and Roberto~Turolla \inst{\ref{in:UniPD},\ref{in:MSSL}}\orcidlink{0000-0003-3977-8760}  
\and Jacco~Vink \inst{\ref{in:Amsterdam}}\orcidlink{0000-0002-4708-4219}  
\and Kinwah~Wu \inst{\ref{in:MSSL}}\orcidlink{0000-0002-7568-8765}  
\and Fei~Xie \inst{\ref{in:GSU},\ref{in:INAF-IAPS}}\orcidlink{0000-0002-0105-5826}  
}

\institute{
Department of Physics and Astronomy, FI-20014 University of Turku,  Finland \label{in:UTU} 
\email{juri.poutanen@utu.fi}
\and 
Institut f\"ur Astronomie und Astrophysik, Universit\"at T\"ubingen, Sand 1, D-72076 T\"ubingen, Germany \label{in:Tub} 
\and 
University of Alabama in Huntsville, NSSTC, Huntsville, AL 35805, USA \label{in:UAH}
\and NASA Marshall Space Flight Center, Huntsville, AL 35812, USA \label{in:NASA-MSFC}
\and 
Institute of Astrophysics, Foundation for Research and Technology-Hellas, GR-70013 Heraklion, Greece \label{in:AstroCrete} 
\and  
Department of Physics, University of Crete, GR-70013, Heraklion, Greece \label{in:PhysCrete} 
\and  INAF Istituto di Astrofisica e Planetologia Spaziali, Via del Fosso del Cavaliere 100, 00133 Roma, Italy \label{in:INAF-IAPS} 
\and 
University of British Columbia, Vancouver, BC V6T 1Z4, Canada \label{in:UBC}
\and
Dipartimento di Fisica, Universit\`{a} degli Studi di Roma ``Tor Vergata'', Via della Ricerca Scientifica 1, 00133 Roma, Italy \label{in:UniRoma2}
\and 
Dipartimento di Fisica, Universit\`{a} degli Studi di Roma ``La Sapienza'', Piazzale Aldo Moro 5, 00185 Roma, Italy \label{in:LaSapienza} 
\and Astrophysics, Department of Physics, University of Oxford, Denys Wilkinson Building, Keble Road, Oxford OX1 3RH, UK \label{in:Oxford}
\and 
Department of Astronomy and Astrophysics, Pennsylvania State University, University Park, PA 16801, USA \label{in:PSU}
\and 
Department of Astrophysics, St. Petersburg State University, Universitetsky pr. 28, Petrodvoretz, 198504 St. Petersburg, Russia \label{in:SPBU}
\and Space Research Institute of the Russian Academy of Sciences, Profsoyuznaya Str. 84/32, Moscow 117997, Russia \label{in:IKI}
\and Nordita, KTH Royal Institute of Technology and Stockholm University, Hannes Alfv\'{e}ns v\"{a}g 12, SE-10691 Stockholm, Sweden  \label{in:Nordita} 
\and 
Mullard Space Science Laboratory, University College London, Holmbury St Mary, Dorking, Surrey RH5 6NT, UK \label{in:MSSL}
\and Istituto Ricerche Solari Aldo e Cele Dacc\`o (IRSOL), Faculty of Informatics, Universit\`a della Svizzera italiana, 6605 Locarno, Switzerland  \label{in:IRSOL} 
\and  Institut f\"ur Sonnenphysik (KIS), Georges-K\"ohler-Allee 401a, 79110 Freiburg, Germany \label{in:KIS} 
\and  Graduate School of Sciences, Tohoku University, Aoba-ku,  980-8578 Sendai, Japan \label{in:Tohoku} 
\and 
Instituto de Astrof\'{i}sica de Andaluc\'{i}a -- CSIC, Glorieta de la Astronom\'{i}a s/n, 18008 Granada, Spain \label{in:CSIC-IAA}
\and 
INAF Osservatorio Astronomico di Roma, Via Frascati 33, 00078 Monte Porzio Catone (RM), Italy \label{in:INAF-OAR}  
\and 
Space Science Data Center, Agenzia Spaziale Italiana, Via del Politecnico snc, 00133 Roma, Italy \label{in:ASI-SSDC}
 \and
INAF Osservatorio Astronomico di Cagliari, Via della Scienza 5, 09047 Selargius (CA), Italy  \label{in:INAF-OAC}
\and 
Istituto Nazionale di Fisica Nucleare, Sezione di Pisa, Largo B. Pontecorvo 3, 56127 Pisa, Italy \label{in:INFN-PI}
\and  
Dipartimento di Fisica, Universit\`{a} di Pisa, Largo B. Pontecorvo 3, 56127 Pisa, Italy \label{in:UniPI} 
\and 
Dipartimento di Matematica e Fisica, Universit\`a degli Studi Roma Tre, via della Vasca Navale 84, 00146 Roma, Italy  \label{in:UniRoma3}
\and  
Istituto Nazionale di Fisica Nucleare, Sezione di Torino, Via Pietro Giuria 1, 10125 Torino, Italy  \label{in:INFN-TO}      
\and  
Dipartimento di Fisica, Universit\`{a} degli Studi di Torino, Via Pietro Giuria 1, 10125 Torino, Italy \label{in:UniTO} 
\and   
INAF Osservatorio Astrofisico di Arcetri, Largo Enrico Fermi 5, 50125 Firenze, Italy 
\label{in:INAF-Arcetri} 
\and  
Dipartimento di Fisica e Astronomia, Universit\`{a} degli Studi di Firenze, Via Sansone 1, 50019 Sesto  Fiorentino (FI), Italy \label{in:UniFI} 
\and   
Istituto Nazionale di Fisica Nucleare, Sezione di Firenze, Via Sansone 1, 50019 Sesto Fiorentino (FI), Italy \label{in:INFN-FI}
\and 
Agenzia Spaziale Italiana, Via del Politecnico snc, 00133 Roma, Italy \label{in:ASI}
\and 
Science and Technology Institute, Universities Space Research Association, Huntsville, AL 35805, USA \label{in:USRA-MSFC}
\and 
Istituto Nazionale di Fisica Nucleare, Sezione di Roma ``Tor Vergata'', Via della Ricerca Scientifica 1, 00133 Roma, Italy  
 \label{in:INFN-Roma2} 
 \newpage
\and 
Department of Physics and Kavli Institute for Particle Astrophysics and Cosmology, Stanford University, Stanford, California 94305, USA  \label{in:Stanford}
\and 
Astronomical Institute of the Czech Academy of Sciences, Boční II 1401/1, 14100 Praha 4, Czech Republic \label{in:CAS-ASU}
\and 
RIKEN Cluster for Pioneering Research, 2-1 Hirosawa, Wako, Saitama 351-0198, Japan \label{in:RIKEN}
\and 
X-ray Astrophysics Laboratory, NASA Goddard Space Flight Center, Greenbelt, MD 20771, USA \label{in:GSFC}
\and 
Yamagata University,1-4-12 Kojirakawa-machi, Yamagata-shi 990-8560, Japan \label{in:Yamagata}
\and 
Osaka University, 1-1 Yamadaoka, Suita, Osaka 565-0871, Japan \label{in:Osaka}
\and 
International Center for Hadron Astrophysics, Chiba University, Chiba 263-8522, Japan \label{in:Chiba}
\and
Institute for Astrophysical Research, Boston University, 725 Commonwealth Avenue, Boston, MA 02215, USA \label{in:BU}  
\and 
Department of Physics and Astronomy and Space Science Center, University of New Hampshire, Durham, NH 03824, USA \label{in:UNH} 
\and 
Istituto Nazionale di Fisica Nucleare, Sezione di Napoli, Strada Comunale Cinthia, 80126 Napoli, Italy \label{INFN-NA}
\and 
Universit\'{e} de Strasbourg, CNRS, Observatoire Astronomique de Strasbourg, UMR 7550, 67000 Strasbourg, France \label{in:Strasbourg}
\and 
MIT Kavli Institute for Astrophysics and Space Research, Massachusetts Institute of Technology, 77 Massachusetts Avenue, Cambridge, MA 02139, USA \label{in:MIT}
\and 
Graduate School of Science, Division of Particle and Astrophysical Science, Nagoya University, Furo-cho, Chikusa-ku, Nagoya, Aichi 464-8602, Japan \label{in:Nagoya}
\and 
Hiroshima Astrophysical Science Center, Hiroshima University, 1-3-1 Kagamiyama, Higashi-Hiroshima, Hiroshima 739-8526, Japan \label{in:Hiroshima}
\and 
Department of Physics and Astronomy, Louisiana State University, Baton Rouge, LA 70803, USA \label{in:LSU} 
\and 
Department of Physics, University of Hong Kong, Pokfulam, Hong Kong \label{in:HKU}
\and 
Universit\'{e} Grenoble Alpes, CNRS, IPAG, 38000 Grenoble, France \label{in:Grenoble}
\and 
Center for Astrophysics, Harvard \& Smithsonian, 60 Garden St, Cambridge, MA 02138, USA \label{in:CfA} 
\and 
INAF Osservatorio Astronomico di Brera, via E. Bianchi 46, 23807 Merate (LC), Italy \label{in:INAF-OAB}
\and 
Dipartimento di Fisica e Astronomia, Universit\`{a} degli Studi di Padova, Via Marzolo 8, 35131 Padova, Italy \label{in:UniPD}
\and 
Anton Pannekoek Institute for Astronomy \& GRAPPA, University of Amsterdam, Science Park 904, 1098 XH Amsterdam, The Netherlands  \label{in:Amsterdam}
\and 
Guangxi Key Laboratory for Relativistic Astrophysics, School of Physical Science and Technology, Guangxi University, Nanning 530004, China \label{in:GSU}
}


\abstract
{Discovery of pulsations from a number of ultra-luminous X-ray (ULX) sources proved that accretion onto neutron stars can produce luminosities exceeding the Eddington limit by several orders of magnitude. 
The conditions necessary to achieve such  high luminosities as well as the exact geometry of the accretion flow in the neutron star vicinity are, however, a matter of debate. 
The pulse phase-resolved polarization measurements that became possible with the launch of the {Imaging X-ray Polarimetry Explorer} (IXPE) can be used to determine the pulsar geometry and its orientation relative to the orbital plane. 
They provide an avenue to test different theoretical models of ULX pulsars.   
In this paper we present the results of three IXPE observations of the first Galactic ULX pulsar Swift~J0243.6+6124  during its 2023 outburst. 
We find strong variations in the polarization characteristics with the pulsar phase. 
The average polarization degree increases from about 5\% to 15\% as the flux dropped by a factor of three in the course of the outburst. 
The polarization angle (PA) as a function of the pulsar phase shows two peaks in the first two observations, but changes to a characteristic sawtooth pattern in the remaining data set. 
This is not consistent with a simple rotating vector model. 
Assuming the existence of an additional constant polarized component, we were able to fit the three observations with a common rotating vector model and obtain constraints on the pulsar geometry. 
In particular, we find the pulsar angular momentum inclination with respect to the line of sight of $i_{\rm p}=$15\degr--\,40\degr, the magnetic obliquity of $\theta_{\rm p} =$60\degr--\,80\degr, and the pulsar spin position angle of $\chi_{\rm p}\approx-50\degr$, which significantly differs from the constant component PA of about 10\degr.   
Combining these X-ray measurements with the optical PA, we find evidence for at least a 30\degr\ misalignment between the pulsar angular momentum and the binary orbital axis.}


\maketitle



\section{Introduction}

Ultra-luminous X-ray sources (ULXs) are non-nuclear objects found in external galaxies and exhibiting high apparent X-ray luminosities exceeding the Eddington limit for stellar-mass black holes \citep[see, e.g.,][for reviews]{Kaaret2017, King2023}. 
A subset of these ULXs has been identified as X-ray pulsars, systems with highly magnetized neutron stars (NSs) undergoing accretion from a companion star 
\citep{Bachetti2014,Furst2016,Israel2017,Carpano2018,RodriguezCastillo2020}. 

The accretion flow in the vicinity of the NS surface is governed by the magnetic field, which channels the accreting matter toward the magnetic poles. 
At low accretion rates, the gas decelerates at the NS surface forming hotspots that radiate in the X-ray band \citep{Zeldovich69}. 
At high accretion rates, the radiation-dominated shock stops the gas above the NS surface forming accretion columns \citep{Basko76,Lyubarskii88,Mushtukov15}. 
The ULX pulsars are characterized by luminosities exceeding the Eddington limit by a factor of 10--500. 
It is possible to achieve such a high luminosity  thanks to a reduction in the interaction cross section between matter and radiation in a strong magnetic field \citep{Mushtukov15}, possibly dominated by quadrupole component   
\citep{Tsygankov2018,Brice21}. 
Alternatively, a strong beaming of radiation by disk outflows can be responsible for high apparent luminosities similarly to ULXs hosting black holes \citep{King01,Poutanen07}. 
However, the latter scenario meets obstacles in the case of strongly magnetized NSs, because the size of the magnetosphere exceeds the size of the  NS by two orders of magnitude, reducing the strength of the outflow \citep{Lipunov82,Chashkina19}. 
Furthermore, the observed radiation cannot be strongly beamed because of the large pulsation amplitude  \citep{Mushtukov21,Mushtukov23b}, implying enormous accretion rates $\gtrsim 10^{19}$\,g\,s$^{-1}$. 
 
Complexities in ULX pulsar studies arise from their considerable distances in external galaxies. 
Discovery of the first Galactic ULX pulsar Swift~J0243.6+6124 (hereafter J0243) in 2017 opened new exciting possibilities.
J0243 was detected by the {\it Swift}/BAT \citep{Cenko2017} and soon after pulsations at a period of 9.86~s were found with the {\it Swift}/XRT \citep{Kennea2017}. 
\citet{Kong2022} reported the discovery of a cyclotron line scattering feature at   $\approx$130\,keV, which implies a strong surface magnetic field of $B\approx 1.5\times 10^{13}$\,G. 
An optical counterpart, a late Oe-type or early Be-type star, was identified as USNO-B1.0 1514+0083050 based on positional coincidence with the {\it Swift}/XRT source \citep{Kennea2017,Bikmaev2017,Kouroubatzakis2017}. 
This star appears in the \textit{Gaia} Data Release 3 (DR3) catalog (ID: 465628193526364416) at the distance determined via parallax of 5.2~kpc\footnote{This is closer than the distance of 6.8~kpc from the \textit{Gaia} DR2 catalog \citep{BJ18} and used in some earlier literature.} \citep{BailerJones2021}. 
Using this distance, the peak luminosity in the 2017 outburst was $\sim 2.5 \times 10^{39}$\,erg \,s$^{-1}$ \citep{Doroshenko2018,Tsygankov2018,WilsonHodge2018}.

Pulsar radiation was predicted to be strongly polarized \citep{Meszaros88}. 
Testing this prediction recently became possible  thanks to the {Imaging X-ray Polarimetry Explorer} (IXPE) that allows us to measure X-ray polarization in the 2--8~keV energy band. 
IXPE discovered pulse-phase dependent polarization in a number of X-ray pulsars  
\citep[e.g.,][see \citealt{Poutanen2024} for a review]{Doroshenko22,Doroshenko23,Tsygankov22,Tsygankov2023,Forsblom2023,Forsblom2024,Malacaria23,Suleimanov2023,Mushtukov23,Heyl24}. 
Variations in the polarization properties with the pulsar phase allowed us to constrain the pulsar geometry.

During the peak of the outburst, the spectrum of J0243 was dominated by the Compton reflection component with a strong iron line associated with the reflection of the primary pulsar radiation from the well formed by the inner edge of the geometrically thick super-Eddington accretion disk \citep{Bykov22}. 
This interpretation was later supported by the detection of the pulsations in the iron line with the \textit{Insight-HXMT} \citep{Xiao24}. 
The reflection component is also expected to be strongly polarized \citep{Matt93,PNS96}, providing information about the reflector geometry and its orientation with respect to the observer.  
A strongly polarized X-ray continuum attributed to reflection was observed by IXPE in other types of compact X-ray sources: in two Seyfert~2 galaxies (the Circinus galaxy, \citealt{Ursini23}; \mbox{NGC 1068}, \citealt{Marin24}) and  in the X-ray binary \mbox{Cyg X-3} \citep{Veledina24}.  
In all these cases the reflection is likely coming from a torus-like structure blocking the direct view  toward the central X-ray source.

\begin{figure}
\centering
\includegraphics[width=0.85\columnwidth]{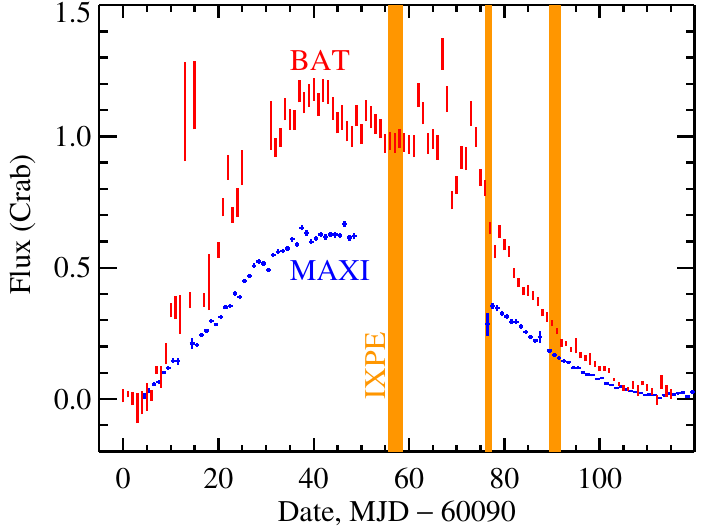}
\caption{Light curve of J0243 in the 2--20 and 15--50~keV energy bands obtained with the MAXI and 
{\it Swift}/BAT monitors, respectively. 
The vertical orange stripes show the times of IXPE observations. 
}
\label{fig:lc}
\end{figure}

In 2023, J0243 underwent another outburst detected first with MAXI on 2023 April~8  \citep{Setoguchi2023}. A few days later the source was localized to be J0243 with the {\it Swift}/XRT \citep{Kennea2023}, and subsequently 9.8~s pulsations were detected with NICER \citep{Ng2023}. The source continued to brighten, reaching peak flux in early July. The light curves of J0243 as seen by the MAXI\footnote{\url{http://maxi.riken.jp/}}  \citep{Matsuoka2009} in the 2--20 keV energy band and \textit{Swift}/BAT\footnote{\url{https://swift.gsfc.nasa.gov/results/transients/}} \citep{Gehrels04} in the 15--50 keV band during the 2023 outburst are shown in Fig.~\ref{fig:lc}.

In this paper we present the results of IXPE observations of J0243 together with accompanying optical polarimetric observations. 
The details of the observations and their analysis are given in Sect.~\ref{sec:Obs}. 
Constraints on the pulsar geometry are worked through in Sect.~\ref{sec:geom}. 
We discuss the results in Sect.~\ref{sec:disc}. 
We conclude with a summary of our findings in Sect.~\ref{sec:summary}.

\section{Observations and data reduction}
\label{sec:Obs}

\subsection{IXPE observations}

The {Imaging
X-ray Polarimetry Explorer} is a NASA mission in partnership with the Italian Space Agency \citep[see a detailed description in][]{Weisskopf2022}, launched by a Falcon 9 rocket on 2021 December 9. 
There are three grazing incidence telescopes on board the observatory. 
Each telescope comprises an X-ray mirror  assembly and a polarization-sensitive detector unit (DU) equipped with a gas-pixel detector 
\citep[GPD;][]{2021AJ....162..208S,2021APh...13302628B}.
These instruments provide imaging polarimetry in the 2--8 keV energy band with a time resolution better than 10\,$\mu$s. 

IXPE observed J0243 three times during 2023 July 20--August 25 (ObsID 02250799), on July 20--22, August 10--11, and August 23--25, with the total exposures of $\simeq$168, 77, and 131~ks per telescope, respectively (see Fig.~\ref{fig:lc}). 
We refer to these observations as Obs. 1, 2, and 3, respectively. 
The data were processed with the \textsc{ixpeobssim} package version 31.0.1 \citep{Baldini22} using the CalDB version 20230702:v13.
Source photons were collected in a circular region with a radius $R_{\rm src}=70\arcsec$ centered at the J0243 position. 
Because of the source brightness, the background was not subtracted following the recommendation by \citet{Di_Marco_2023}. 
We performed the unweighted analysis (i.e.,  all events are taken into account independently of the quality of track reconstruction; \citealt{Di_Marco_2022}) of the IXPE data.

\begin{table}
\centering
\caption{Orbital parameters of J0243 from the \textit{Fermi}/GBM data. }
\begin{tabular}{lcc}
\hline
\hline
Epoch of 90\degr\ mean longitude & $T_{\pi/2}$ (JED)  & 2458116.0970 \\
Orbital period &  $P_{\rm orb}$ (d)  & 27.698899 \\
Period derivative & $\dot{P}_{\rm orb}$ (d\,d$^{-1}$)  & 0.000000 \\
Projected semimajor axis & $a_{\rm x}\sin i$ (lt-s) & 115.531 \\ 
Longitude of periastron
& $\omega$ (deg) & $-74.05$ \\ 
Eccentricity & $e$ & 0.1029 \\ 
\hline
\end{tabular}
\label{tab:orbeph}
\end{table}

\begin{table}
\centering
\caption{Timing parameters used to fold IXPE data.}
    \begin{tabular}{cccc}
    \hline\hline
       Segment  & Epoch (MET) & $P$ (s) \tablefootmark{a} & 
       $\dot{P}$ ($10^{-9}$\,s\,s$^{-1}$) \tablefootmark{b} \\
       \hline
         1 & 206636790.164 &  9.795773(3) & $-1.67(3)$ \\
         2 & 208396144.591 &  9.79308(1) & $-1.2(1)$ \\
         3 & 209558782.987 &  9.792140(5) & $-2.75(6)$ \\
         \hline
    \end{tabular}
\tablefoot{Folding epoch is fixed to first pulse arrival time; uncertainties are quoted at the  1$\sigma$ confidence level.
\tablefoottext{a}{Pulse period.}
\tablefoottext{b}{Pulse period derivative.}
}
    \label{tab:spin}
\end{table}

\begin{table*}[]
\centering
\caption{Spectral parameters for the best-fit model obtained with \textsc{xspec} for Observations 1, 2, and 3.}
\begin{tabular}{llcccc}
\hline \hline
Component & Parameter & Unit & Obs. 1 & Obs. 2 & Obs. 3 \\
\hline
\texttt{tbabs} & $N_{\mathrm{H}}$ & $10^{22}\mathrm{\;cm^{-2}}$ & $1.60\pm0.09$ & $1.23\pm0.17$ & $0.88\pm0.20$ \\
\texttt{bbody} & $kT_{\rm bb}$ & keV      & $0.52\pm0.02$ & $0.60\pm0.04$ & $0.92\pm0.03$ \\
& normalization &     & $0.015\pm0.001$ & $0.008\pm0.001$ & $0.005\pm0.001$ \\
\texttt{powerlaw} & Photon index &  & $1.25\pm0.02$ & $0.92\pm0.05$ & $0.53\pm0.12$ \\
& normalization &     & $1.04\pm0.04$ & $0.37\pm0.04$ & $0.08\pm0.03$ \\
\texttt{constant} & DU2 &  & $1.032\pm0.001$ & $1.036\pm0.002$ & $1.026\pm0.002$ \\
 & DU3 &  & $1.015\pm0.001$ & $1.018\pm0.002$ & $1.010\pm0.002$ \\
\hline
 &  $F_{2-8\;{\rm keV}}$ \tablefootmark{a}& $\mathrm{10^{-9}\;erg\;cm^{-2}\;s^{-1}}$  & $6.7\pm1.0$ & $4.1\pm0.6$ & $1.85\pm0.25$ \\
 & $L_{2-8\;{\rm keV}}$\tablefootmark{b} & $10^{37}$\,erg\;s$^{-1}$  & 2.4 & 1.4 & 0.6 \\
 & $L_{\rm bol}$\tablefootmark{c} & $10^{37}$\,erg\;s$^{-1}$  & 9.1 & 5.3 & 2.3 \\
 & $\chi^2$/d.o.f. &  & 677.4/437 & 540.1/437 & 503.8/437 \\
\hline  
\end{tabular}
\tablefoot{Uncertainties are given at the 68.3\% (1$\sigma$) confidence level and were obtained using the \texttt{error} command in \textsc{xspec} with $\Delta\chi^2=1$ for one parameter of interest. 
\tablefoottext{a}{Observed flux in the 2--8 keV range.}
\tablefoottext{b}{Unabsorbed luminosity for the assumed distance of $d=5.2$~kpc.}
\tablefoottext{c}{Total luminosity in the range 0.5--100~keV assuming a bolometric correction factor of 3.8.}}
\label{tab:best-fit}
\end{table*}

\subsection{Timing}

The event arrival times were corrected to the Solar System barycenter using the standard \texttt{barycorr} tool from the \textsc{ftools}\footnote{\url{http://heasarc.gsfc.nasa.gov/ftools}} package \citep{Blackburn95}  and accounting for the effects of binary motion using the orbital parameters from the \textit{Fermi}/GBM\footnote{\url{https://gammaray.nsstc.nasa.gov/gbm/science/pulsars.html}} (see Table~\ref{tab:orbeph}).
The spin period (of $P\approx9.79$\,s) and the spin period derivative were then determined for each IXPE observation segment using an epoch-folding search method, and then refined using phase-connection technique, which allows the phase of each event within a given segment to be unambiguously defined. 
The results are presented in Table~\ref{tab:spin}. 
However, the gaps between the individual observation segments coupled to the complex evolution of intrinsic spin frequency and pulse profiles together with the  remaining uncertainties in orbital parameters preclude the direct connection of phases between the segments. 

\begin{figure}
\centering
\includegraphics[width=0.90\columnwidth, clip]{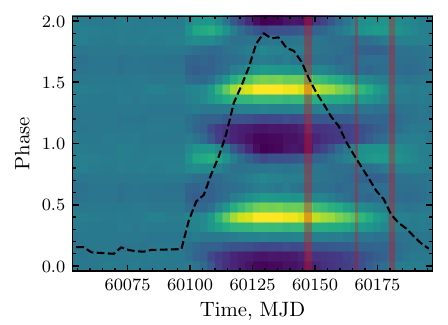} 
\includegraphics[width=0.85\columnwidth]{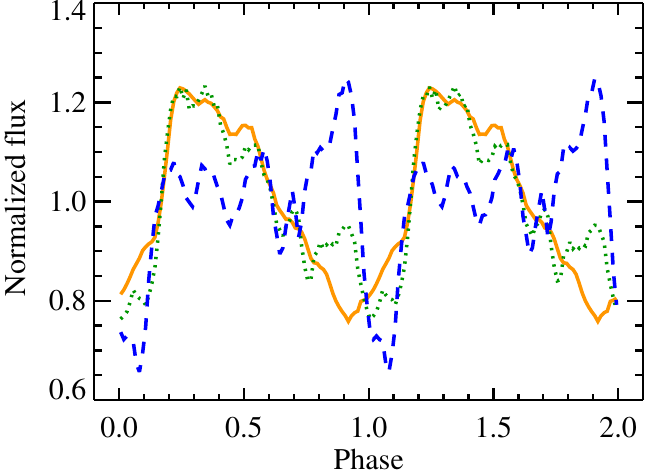}
\caption{Evolution of the pulse profiles during the outburst. 
Top: Color-coded pulse profiles as observed by \textit{Fermi}/GBM in units of relative intensities \citep[see Appendix~A.2 in][]{WilsonHodge2018};    yellow  corresponds to the maxima of the pulse and dark blue to the minima. 
The black dashed line shows the evolution of the pulsed flux in the 12--50~keV range during the outburst as seen by \textit{Fermi}/GBM\protect\footnotemark{} with the peak flux being $4.8\times10^{-9}$\,erg\,cm$^{-2}$\,s$^{-1}$.
The shaded vertical stripes mark the times of IXPE observations. 
Bottom: Normalized pulse profiles in the 2--8 keV band as observed by IXPE in three observations and shown with the solid orange, dotted green, and dashed blue lines for Obs. 1, 2, and 3, respectively.}
\label{fig:pulse-profiles}
\end{figure}

The absolute phase alignment between the segments was therefore determined independently. 
Inspection of the pulse profile shape observed by IXPE (see bottom panel of Fig.~\ref{fig:pulse-profiles}) reveals some common features:  the minimum around phase zero, several  subpeaks within the main peak, and a peak around phase 0.9 appearing late in the outburst. 
The observed phases of these features can thus be used to determine absolute phase offset for each segment such that all peaks and dips
(determined through the fitting of Gaussian functions) appear at the same pulse phase. 
The residual scatter can then be used to assess the remaining uncertainty in the final phase alignment, which we estimate to be below 1\%. 
The phase alignment can also be checked through the comparison of the observed IXPE pulse profiles with hard X-ray pulse profiles observed by \textit{Fermi}/GBM. 

To this end, we used the enhanced products provided by the GBM team containing Fourier coefficients of the pulse profiles for a set of time intervals with typical duration of $\sim$3\,d. 
These are expected to change smoothly, and since \textit{Fermi} data contains no data gaps, the individual pulse profiles can be aligned through the cross-correlation of subsequent time intervals. 
The resulting pulse profile evolution is presented in the top panel of Fig.~\ref{fig:pulse-profiles}. 
We see that changes similar to those revealed by IXPE  also occur in the GBM data:  the secondary peak around phase 0.9 appears to be present only at lower luminosities and disappears at higher luminosities (i.e., our initial alignment using IXPE data alone proves to be robust).

\begin{figure}
\centering
\includegraphics[width=0.9\linewidth]{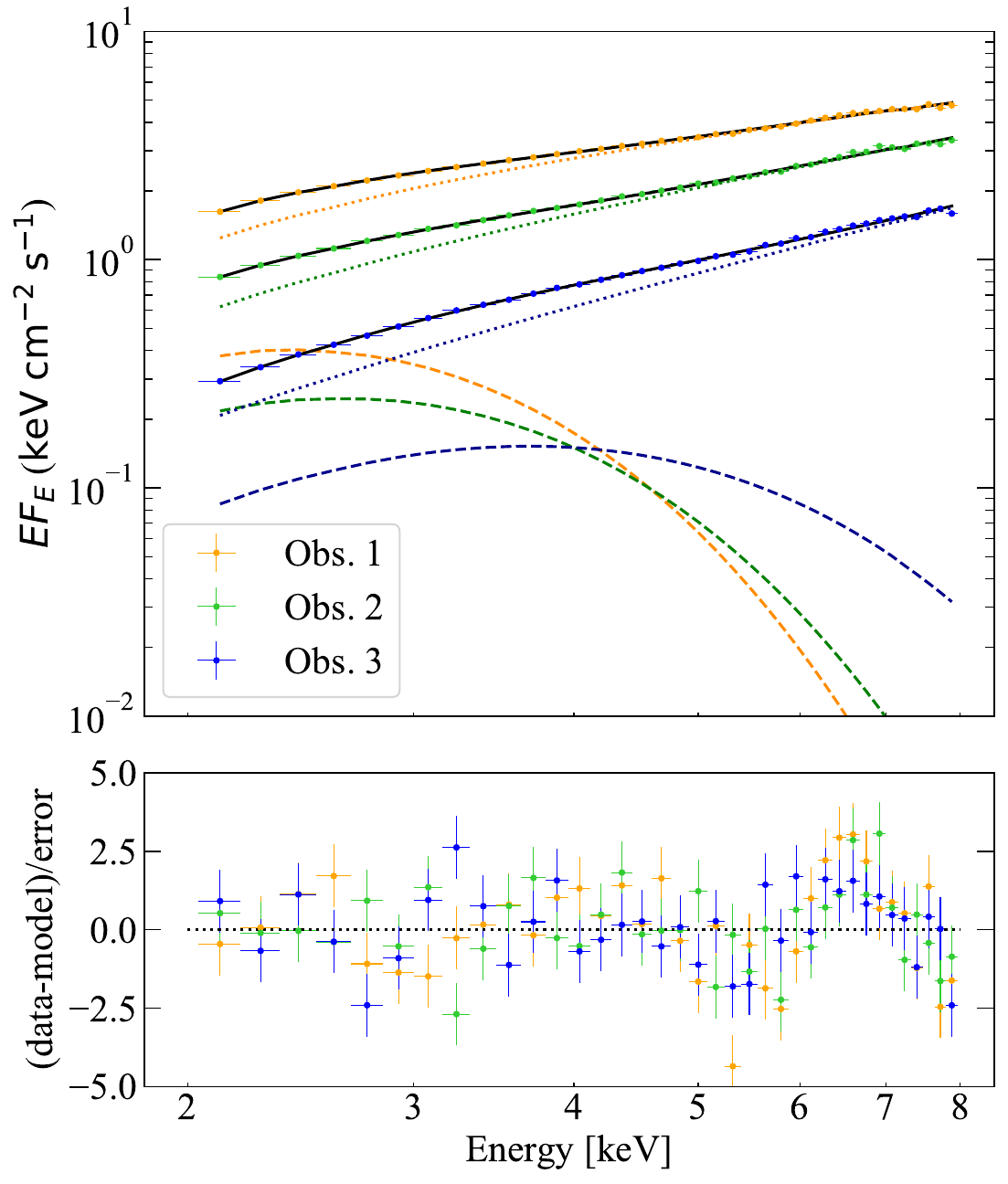}
\caption{IXPE spectra of J0243  in $EF_E$ representation. 
The red, green, and blue symbols and lines correspond to Obs. 1, 2, and 3, respectively. 
The solid, dashed, and dotted lines show the total model spectrum, the blackbody, and the power law, respectively. 
The bottom panel shows the fit residuals. } 
\label{fig:spec}
\end{figure}

\footnotetext{\url{https://gammaray.nsstc.nasa.gov/gbm/science/pulsars/lightcurves/swiftj0243.html}} 

\subsection{Spectral analysis}

For the spectral analysis, Stokes $I$ spectra for Obs. 1, 2, and 3 were extracted using the \texttt{xpbin} tool’s \texttt{PHA1} algorithm in \textsc{\mbox{ixpeobbsim}}, giving a set of three spectra (one for each DU) per observation.
The energy spectra were binned to have at least 30 counts per energy channel.
The spectra corresponding to the individual observations were fitted separately with the \textsc{xspec} package version 12.14.0 \citep{Arn96} using $\chi^2$ statistics.
The uncertainties are given at the 68.3\% confidence level.
Considering the energy resolution and energy range of IXPE, we used a simplified model consisting of an absorbed power law plus a blackbody component to fit the J0243 spectra.
To account for the interstellar absorption, a multiplicative component \texttt{tbabs} with the abundances from \citet{Wilms2000} was applied. 
A re-normalization constant \texttt{const} was introduced to account for possible discrepancies between effective areas of the different DUs and was fixed at unity for DU1.
The full spectral model  is thus
\begin{equation}
\texttt{tbabs$\times$(bbody+powerlaw)$\times$const} . 
\label{eq:fit}
\end{equation}
The results of the spectral fitting are shown in Fig.~\ref{fig:spec}, and the best-fit parameters are given in Table~\ref{tab:best-fit}. 
We see that the fit quality is not good, which is caused partially by systematic errors in the instrument response and partially by the presence of a weak iron line around 6.5~keV. 
However, this does not preclude us from measuring the flux in the IXPE band. 
For calculations of the total luminosity we estimated the bolometric correction factor for our IXPE observations from the {\it NuSTAR} spectra at similar luminosities during the 2017 outburst \citep{Tsygankov2018,Bykov22}. 
The maximum bolometric luminosity of the 2023 outburst is $10^{38}$~erg~s$^{-1}$, a factor of 20 lower than the peak of the 2017 outburst (see penultimate row of Table~\ref{tab:best-fit}). 

\begin{figure}
\centering
\includegraphics[width=0.85\linewidth]{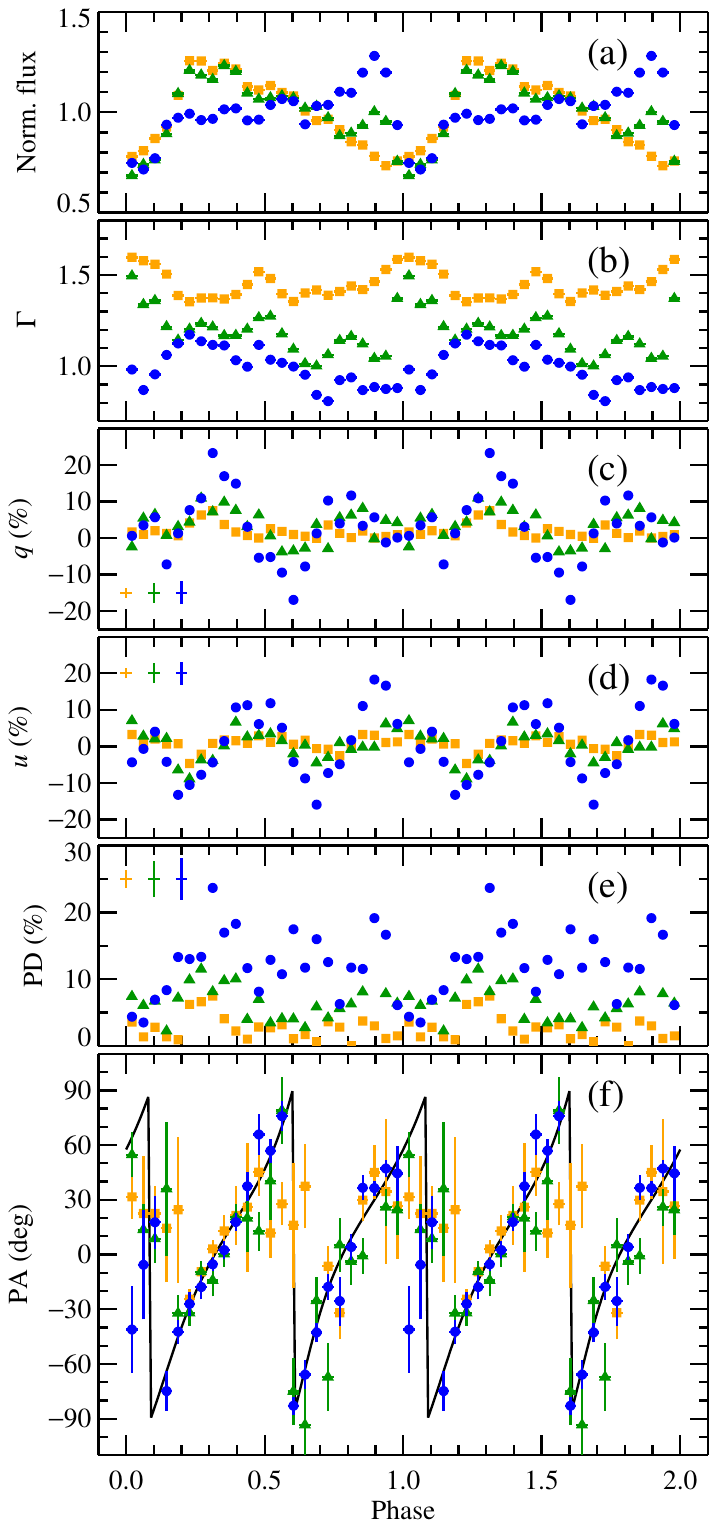}
\caption{Results from the pulse-phase-resolved analysis of J0243 in the 3--8 keV range, combining data from all DUs.  
\textit{Panel (a)}: Pulse profile.
\textit{Panel (b)}: Photon spectral index.  
\textit{Panels (c)} and \textit{(d)}: Dependence of the Stokes $q$ and $u$ parameters. 
\textit{Panels (e)} and \textit{(f)}:  PD and PA.
The data from Obs.\,1, 2, and 3 are shown as   orange squares, green triangles, and blue circles, respectively. 
The typical error bar corresponding to 1$\sigma$ uncertainty is shown in panels (c)--(e). 
The black solid curve is the best-fit RVM to the original PA data points (right column of Table~\ref{tab:rvm}). } 
\label{fig:phase-res-pcube}
\end{figure}

We also studied variations in  the spectrum over the pulse phase. 
Because the photon statistics do  not allow us to use complex models, we fitted the phase-resolved spectra with a simple \texttt{powerlaw} model and fixed  the hydrogen column density $N_{\rm H}$ in the \texttt{tbabs} model at the best-fit phase-average values of 1.23, 1.01, and 1.56$\times10^{22}$~cm$^{-2}$ for Obs.~1, 2, and 3, respectively. 
Variations in the best-fit photon index $\Gamma$ with pulse phase are shown in Fig.~\ref{fig:phase-res-pcube}b.

\begin{figure}
\centering
\includegraphics[width=0.75\columnwidth]{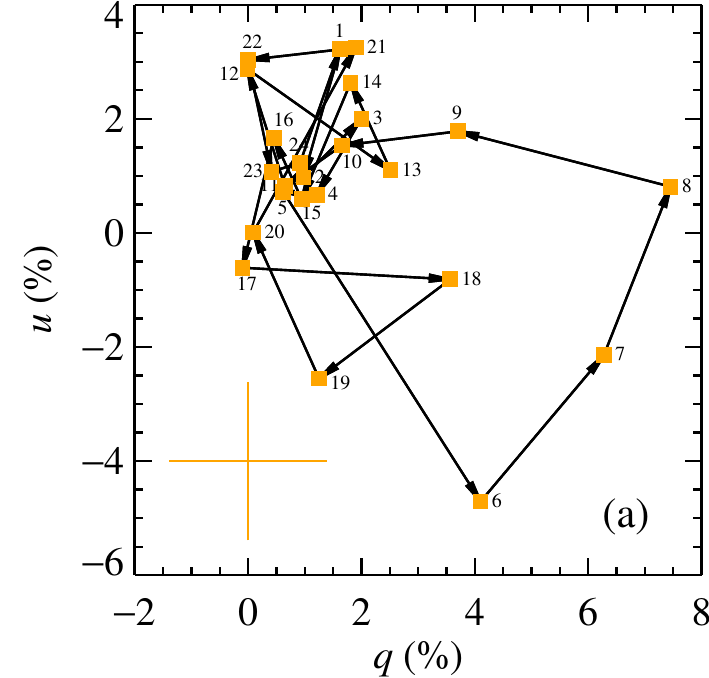}
\includegraphics[width=0.75\columnwidth]{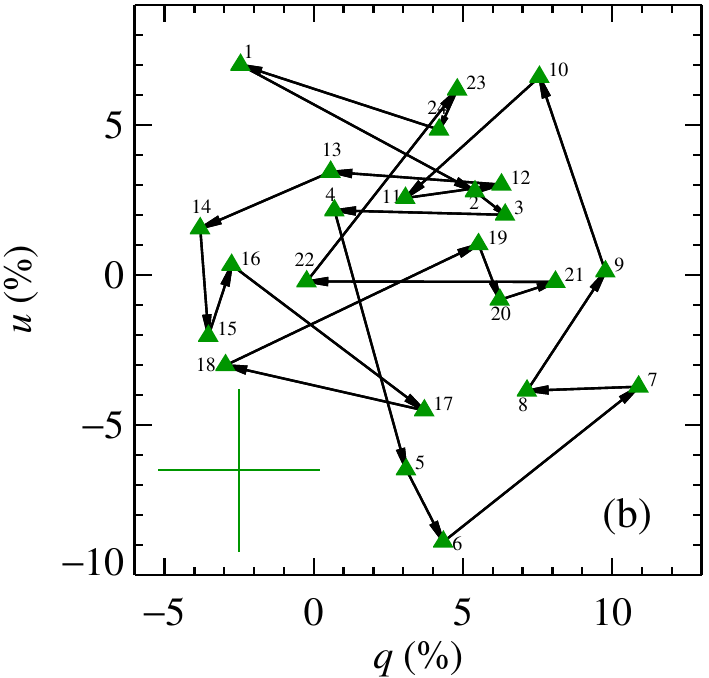}
\includegraphics[width=0.75\columnwidth]{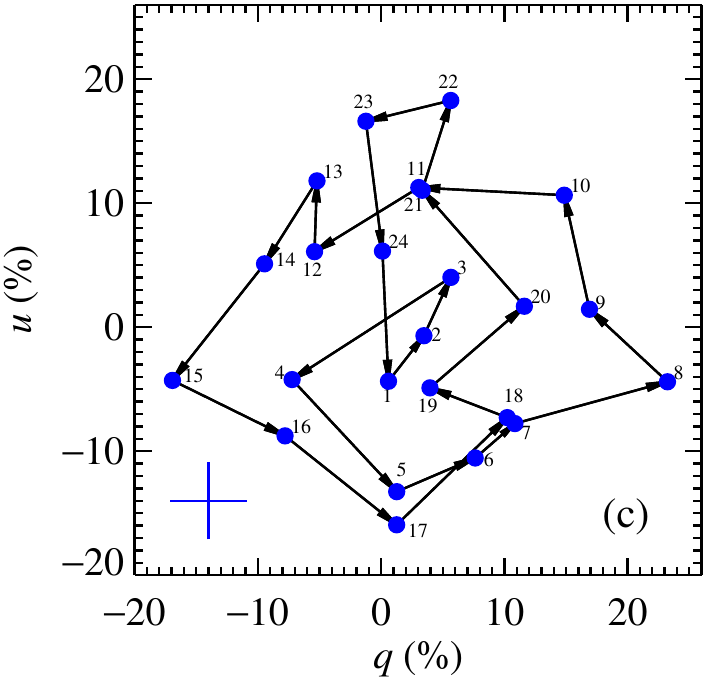}
\caption{Normalized Stokes parameters $q$ and $u$ for the phase-resolved polarimetric analysis using \texttt{pcube} (DUs combined), for the 3--8 keV energy band for Obs. 1 (panel a), 2 (panel b), and 3 (panel c). 
The typical error bar corresponding to 1$\sigma$ uncertainty is shown in each panel. 
The phase bins are numbered. The scale in each   panel is different.}
\label{fig:Stokes_resolved}
\end{figure}

\subsection{X-ray polarimetric analysis}

The polarimetric parameters of J0243 were extracted utilizing the \texttt{pcube} algorithm (\texttt{xpbin} tool) in the \textsc{ixpeobssim} package \citep{Baldini22}, which follows the description of \citet{2015-Kislat}. 
We derived the normalized Stokes parameters $q=Q/I$ and $u=U/I$, and subsequently the polarization degree PD=$\sqrt{q^2+u^2}$  and polarization angle PA=$\frac{1}{2}\arctan(u/q)$. 
The uncertainties are given at the 68.3\% confidence level unless stated otherwise.

The data were divided into 24 separate pulse-phase bins. 
The polarization was found to be low in the 2--3 keV band. 
This motivated us to use for the following analysis only the 3--8 keV data to reduce the noise. 
The  pulse-phase dependence of the normalized Stokes parameters $q$ and $u$ is displayed in Fig.~\ref{fig:phase-res-pcube}c and d, and their evolution on the $(q,u)$-plane is shown in Fig.~\ref{fig:Stokes_resolved}. 
The PD (shown in Fig.~\ref{fig:phase-res-pcube}e) grows with time, from less than 8\% in Obs.\,1 to $\sim$12\% in Obs.\,2, and reaches a maximum of 24\% in Obs.\,3.  
The PA (Fig.~\ref{fig:phase-res-pcube}f) has a double peak profile in all three observations. 
In Obs. 1 and 2 the amplitude of variations is about 100\degr,\footnote{This is in strong disagreement with the recent analysis of Obs.~1 by \citet{Majumder24}, who used only five phase bins.} while in Obs. 3 the data are consistent with two full revolution by 180\degr\ during the spin period. 
We do not see an obvious correlation between spectral shape and polarimetric properties.

\subsection{Optical polarization studies}

\begin{figure}
\centering
\includegraphics[width=0.9\columnwidth]{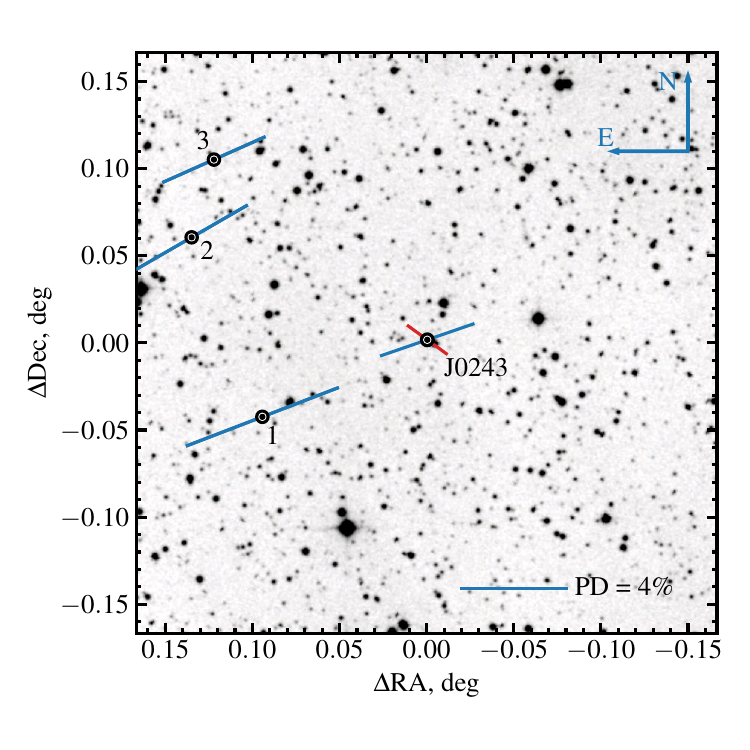}
\caption{Finding chart and the \textit{R}-band polarization of J0243 (in the center) and three nearby field stars, which are situated at a similar distance according to \textit{Gaia} parallaxes. The blue bars show the observed polarization of the source and field stars, while the red bar corresponds to the intrinsic polarization of the source from the Robopol observations, taking star \#2 as an estimate of the interstellar contribution (see Table~\ref{tab:robopol}).}
\label{fig:sky}
\end{figure}

\begin{figure}
\centering
\includegraphics[width=0.9\linewidth]{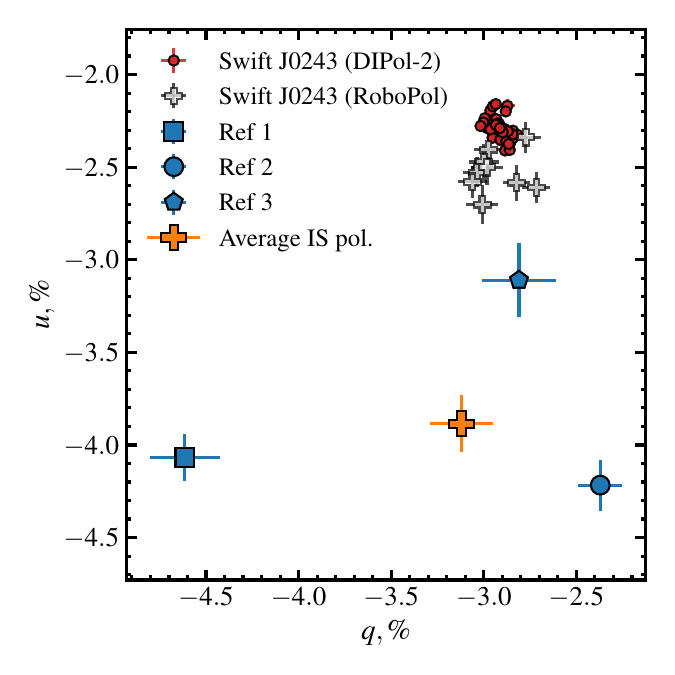}
\caption{Observed normalized Stokes parameters of  the \textit{R}-band optical polarization of J0243 and the three nearby field stars shown in Fig.~\ref{fig:sky}.}
\label{fig:quplane}
\end{figure}

\begin{table}
\caption{Optical polarization of J0243 as observed with Robopol in 2023 in \textit{R}-band.} 
\centering
\begin{tabular}{lcc}
\hline
\hline
\multicolumn{3}{c}{Observed polarization of J0243} \\  
 HJD  &  $q$ (\%) & $u$ (\%) \\
 \hline
2460153.5530 & $-2.97\pm0.08$ & $-2.41\pm0.08$ \\    
2460154.5744 &  $-3.00\pm0.08$ & $-2.47\pm0.10$ \\ 
2460155.5594 &   $-2.77\pm0.08$ & $-2.34\pm0.08$ \\ 
2460156.5468 &    $-3.04\pm0.07$ & $-2.53\pm0.08$ \\ 
2460157.5855 &   $-3.06\pm0.08$ & $-2.58\pm0.09$ \\ 
2460162.5357 &   $-3.00\pm0.08$ & $-2.70\pm0.10$ \\ 
2460181.5359 &   $-2.98\pm0.08$ & $-2.50\pm0.10$ \\ 
2460182.6066 &   $-2.72\pm0.08$ & $-2.61\pm0.08$ \\ 
2460184.5981 &    $-2.82\pm0.07$ & $-2.58\pm0.10$ \\ 
\hline 
Average observed  & $-2.93 \pm 0.03$ & $-2.52 \pm 0.03$ \\ \hline
\multicolumn{3}{c}{Interstellar polarization} \\  
Star \#1 & $-4.61\pm0.19$ & $-4.07\pm0.13$ \\
Star \#2 & $-2.37\pm0.12$ & $-4.22\pm0.14$ \\ 
\hline
\multicolumn{3}{c}{Intrinsic polarization} \\  
IS=star \#1  & $\phantom{-}1.68 \pm 0.19$ & $\phantom{-}1.55 \pm 0.13$ \\ 
IS=star \#2  & $-0.56 \pm 0.12$ & $\phantom{-}1.70 \pm 0.14$ \\ 
\hline 
  &  PD (\%) & PA (deg) \\
\hline
IS=star \#1 & $\phantom{-}2.29\pm 0.16$ & $21 \pm 2$ \\
IS=star \#2 & $\phantom{-}1.79\pm 0.13$ & $54 \pm 2$ \\
\hline 
\end{tabular}
\tablefoot{Normalized Stokes parameters $q$ and $u$ are presented for the observed optical polarization of the source, the interstellar (IS) polarization, and the intrinsic polarization obtained by subtracting the IS polarization from the observed values. 
The PD and PA $\chi_{\rm o}$ of the intrinsic optical polarization are computed from the intrinsic $q$ and $u$.
Uncertainties are 1$\sigma$. }
\label{tab:robopol}
\end{table}

The IXPE observations were accompanied by optical polarimetric observations in the $R$-band with the Robopol polarimeter located in the focal plane of the 1.3\,m telescope of the Skinakas observatory (Greece).
The observations were performed between 2023 July 28 and August 28 with multiple pointings. 
The data are presented in Table~\ref{tab:robopol}.
We determined the intrinsic source polarization by subtracting the interstellar  polarization using stars \#1 and \#2 (see Fig.~\ref{fig:sky}), which are located $\sim$3\arcmin--4\arcmin\ away from J0243 at a compatible distance of $\sim$5.7~kpc, as given by the \textit{Gaia} EDR3 data \citep{BailerJones2021}.  

We also analyzed optical polarimetric measurements of the source during its previous outburst in 2017 obtained with the Double-Image Polarimeter \citep[DIPol-2;][]{Piirola2014} at the T60 telescope at Haleakala, Hawaii (see Table~\ref{tab:dipol} in Appendix~\ref{sec:app1}). 
On average, 20 measurements of polarization were taken each night. 
We obtained the nightly averaged values using the $2\sigma$ clipping method \citep{Kosenkov2017,Kosenkov2021}.
To estimate the interstellar polarization, we also observed  field star \#3 (see Fig.~\ref{fig:sky}), which was reasonably bright and also has a parallax similar to that of the source.  

The resulting normalized Stokes parameters of J0243 and the field stars are shown in Fig.~\ref{fig:quplane}. 
The intrinsic optical PD lies  between $\sim$1\% and 2.5\% and the intrinsic optical PA $\chi_{\rm o}$ is in the range 20\degr--50\degr, depending on the choice of the field star.

\section{Pulsar geometry} 
\label{sec:geom}

\subsection{RVM} 
\label{sec:rvm}

\begin{figure}
\centering
\includegraphics[width=0.75\columnwidth]{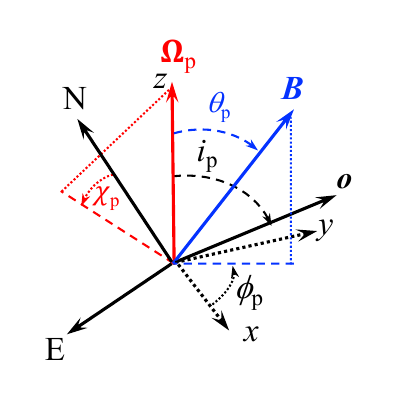}
\caption{Geometry of the pulsar and main parameters of the RVM. 
The pulsar angular momentum $\vec{\Omega}_{\rm p}$ makes an angle $i_{\rm p}$ with respect to the observer direction $\vec{o}$. 
The angle $\theta_{\rm p}$ is the magnetic obliquity, i.e., the angle between magnetic dipole and the rotation axis. 
The pulsar phase $\phi$ is the azimuthal angle of vector $\vec{B}$ in the plane $(x,y)$ perpendicular to  $\vec{\Omega}_{\rm p}$ and  $\phi=\phi_{\rm p}$ when $\vec{B}$, $\vec{\Omega}_{\rm p}$, and $\vec{o}$ are coplanar. 
The pulsar position angle $\chi_{\rm p}$ is the angle measured counterclockwise between the direction to the north (N) and the projection of $\vec{\Omega}_{\rm p}$ on the plane of the sky (NE).}
\label{fig:geometry}
\end{figure}

Previous IXPE data on a number of X-ray pulsars \citep{Doroshenko22,Doroshenko23,Tsygankov22,Tsygankov2023,Mushtukov23}  were well described by the rotating vector model \citep[RVM;][]{Radhakrishnan69,Meszaros88}. 
In this model, the evolution of the PA with pulsar phase is related  to the projection of the magnetic dipole on the plane of the sky. 
If radiation is dominated by the ordinary-mode  (O-mode) photons, the PA $\chi$ is given by Eq.\,(30) in \citet{Poutanen2020}: 
\begin{equation} \label{eq:pa_rvm}
\tan (\chi-\chi_{\rm p})=\frac{-\sin \theta_{\rm p}\ \sin (\phi-\phi_{\rm p})}
{\sin i_{\rm p} \cos \theta_{\rm p}  -  \cos i_{\rm p} \sin \theta_{\rm p}  \cos (\phi-\phi_{\rm p}) } .
\end{equation} 
Here $\chi_{\rm p}$ is the position angle (measured from north to east) of the pulsar angular momentum, $i_{\rm p}$ is the inclination of the pulsar spin with respect to the line of sight, $\theta_{\rm p}$ is the magnetic obliquity (i.e., the angle between the magnetic dipole and the spin axis), and $\phi_{\rm p}$ is the phase at which the northern magnetic pole passes in front of the observer (see Fig.~\ref{fig:geometry} for geometry).

\begin{table*}
\caption{Best-fit RVM parameters for the three IXPE observations.} 
\centering
\begin{tabular}{lcccc}
\hline
\hline 
Parameter  & Obs.~1 & Obs.~2 & Obs.~3 & Obs.~1--3 \\
\hline
$i_{\rm p}$  (deg) & 
$80\pm3$ & $60\pm5$ & $33^{+6}_{-8}$ & $47^{+3}_{-4}$ \\
$\theta_{\rm p}$ (deg)  & $87\pm2$ & $88\pm3$  &  $75^{+4}_{-6}$ & $83\pm2$  \\
$\chi_{\rm p}$ (deg)  & 
$-70\pm4$ & $-87\pm7$ & $-66\pm7$ & $-67\pm4$ \\
$\phi_{\rm p}/(2\pi)$ & $0.70\pm0.01$ & $0.60\pm0.03$  & $0.65\pm0.02$ & $0.64\pm0.01$ \\
AIC\tablefootmark{a} &    &    &    & 93.0 \\
$\chi^2$/dof & 25.1/20 & 83.3/20  & 65.7/20 & 276/68 \\
\hline
\end{tabular}
\tablefoot{Uncertainties on parameters correspond to 
$\Delta \log L=1$ for the log-likelihood function \eqref{eq:logL} and are  equivalent to $1\sigma$. 
\tablefoottext{a}{Akaike information criterion AIC=$2K+\log L$ \citep{Akaike74}, with $K$ being the number of model parameters.}
}
\label{tab:rvm}
\end{table*}

If radiation is dominated by the extraordinary mode (X-mode), the PA is rotated by 90\degr\ with respect to the O-mode.
The general relativistic effects are significant only if the NS rotates at millisecond periods \citep[see][]{Poutanen2020,Loktev20}. 
The RVM is also theoretically justified, because the polarization vector of the radiation produced at the magnetic poles rotates adjusting to the magnetic field geometry until photons reach the adiabatic radius at a few tens of stellar radii due to the vacuum birefringence  \citep{2003MNRAS.342..134H}, where the magnetic field is predominantly dipole (see Sect.~\ref{sec:pdflux}).

We fit the RVM to a given set of Stokes parameters $(q,u)$ as a function of pulsar phase. 
The PA distribution does not conform to a normal distribution, and hence we use the  probability density function from \citet{Naghizadeh1993}, 
 \begin{equation} \label{eq:PA_dist}
 G(\chi) = \frac{1}{\sqrt{\pi}}  \left\{  \frac{1}{\sqrt{\pi}}  +  \eta {\rm e}^{\eta^2}  \left[ 1 + {\rm erf}(\eta) \right] \right\} {\rm e}^{-p_0^2/2} ,
 \end{equation}
where $p_0=\sqrt{q^2+u^2}/\sigma_{\rm p}$ is the measured PD in units of the error,  $\eta=p_0 \cos[2(\chi-\textrm{PA})]/\sqrt{2}$, and \mbox{erf} is the error function. 
Here $\chi$ is the prediction of the RVM given by Eq.~\eqref{eq:pa_rvm} for a given set of parameters and PA=$\frac{1}{2}\arctan(u/q)$. 
The best fit with the RVM to the pulse-phase dependent $(q,u)$  is obtained by minimizing the log-likelihood function 
\begin{equation}
\label{eq:logL}
    \log L= -2 \sum_{i,k} \ln   G(\chi_{i,k}) , 
\end{equation} 
with the sum taken over all phase bins $k$ for a given observation $i=1,2,3$ or summing over all of them.  
The error on a parameter can be obtained by varying that parameter and fitting all other parameters until  $\Delta \log L=1$ is reached. 
We can also, in principle, use a $\chi^2$ statistic to evaluate the quality of the fit. 
This is possible because for a highly significant detection of polarization, the PA is distributed almost normally, while the low-significance data points (when the PA distribution is far from normal) contribute very little to the $\chi^2$.

First, we fit the RVM to individual observations. 
The results are given in Table~\ref{tab:rvm}. 
We find that the best-fit  parameters are vastly different, for example the pulsar inclination for the three observations is $i_{\rm p}=$80\degr, 60\degr, and 33\degr. 
We note that the corresponding $\chi^2$ values for the best fits are 25.1, 83.3, and 65.7 for 20 degrees of freedom (d.o.f.) for Obs.~1--3, respectively.   
Thus, only the fit to Obs.~1 is reasonably good. 
If, on the other hand, we force the RVM parameters to be the same, we get the best-fit parameters of $i_{\rm p}=47\degr$, $\theta_{\rm p}=83\degr$, $\chi_{\rm p}=-67\degr$, and $\phi_{\rm p}=0.64$ (this RVM is shown in Fig.~\ref{fig:phase-res-pcube}f), but an unacceptable fit with $\chi^2/\rm d.o.f.=276/68$.
Thus, we find that the RVM does not provide a good description of the data. 

A similar situation occurred with the brightest transient pulsar observed by IXPE, \mbox{LS V +44 17}/\mbox{RX J0440.9+4431}, which showed the peak luminosity of $4.3\times10^{37}$\,erg\,s$^{-1}$ during its giant outburst in 2023 January--February \citep{Salganik23}. 
In this object, significant changes in the RVM parameters  were detected \citep{Doroshenko23}. 
There, the pulse-phase resolved Stokes  $q$ and $u$ parameters for two epochs of observations traced similar patterns on the $(q,u)$ plane, but the figures were shifted relative to each other \citep[see][]{Poutanen2024}.    
An additional, phase-independent polarized component was introduced to align the results with the RVM predictions. 
We now apply the same idea to the J0243 data.

\begin{table}
\caption{Best-fit parameters of the two-component model.} 
\centering
\begin{tabular}{lccc}
\hline
\hline 
Parameter  & Obs.~1 & Obs.~2 & Obs.~3  \\
\hline
$i_{\rm p}$  (deg) & 
\multicolumn{3}{c}{$40^{+9}_{-11}$}   \\
$\theta_{\rm p}$ (deg)  & \multicolumn{3}{c}{$80^{+4}_{-5}$}    \\
$\chi_{\rm p}$ (deg)  & \multicolumn{3}{c}{$-45^{+16}_{-10}$}   \\
$\phi_{\rm p}/(2\pi)$ & \multicolumn{3}{c}{$0.70^{+0.05}_{-0.03}$}   \\ 
$Q_{\rm c}$ (\%) & $1.3\pm0.4$ & $3.4^{+1.0}_{-0.8}$ & $1.9^{+2.3}_{-1.3}$ \\
$U_{\rm c}$ (\%) & $0.7\pm0.4 $ & $-0.6\pm0.9$ & $-1.6\pm1.7$ \\ 
$\chi_{\rm c}$ (deg) &  $14\pm8$ & $-5\pm8$ & $-20\pm21$ \\
$P_{\rm c}I_{\rm c}$ (\%) & $1.4\pm0.4$ & $3.4\pm0.9$ & $2.5\pm1.8$ \\ 
AIC\tablefootmark{a} &  \multicolumn{3}{c}{ 71.5} \\ 
$\chi^2$/dof & \multicolumn{3}{c}{150/62}   \\
\hline
\multicolumn{4}{c}{Same PA for constant component} \\
$i_{\rm p}$  (deg) & 
\multicolumn{3}{c}{$29^{+9}_{-15}$}   \\
$\theta_{\rm p}$ (deg)  & \multicolumn{3}{c}{$76^{+5}_{-15}$}    \\
$\chi_{\rm p}$ (deg)  & \multicolumn{3}{c}{$-49^{+12}_{-11}$}   \\
$\phi_{\rm p}/(2\pi)$ & \multicolumn{3}{c}{$0.68\pm0.03$}   \\  
$\chi_{\rm c}$ (deg) & \multicolumn{3}{c}{$8\pm7$} \\
$P_{\rm c}I_{\rm c}$ (\%) & $1.5\pm0.4$ & $3.1\pm0.7$ & $1.5\pm1.0$ \\ 
AIC\tablefootmark{a} &  \multicolumn{3}{c}{73.8} \\
$\chi^2$/dof & \multicolumn{3}{c}{156/64} \\  
\hline
\end{tabular}
\tablefoot{Uncertainties on parameters correspond to $\Delta \log L=1$ for the log-likelihood function \eqref{eq:logL} and are  equivalent to $1\sigma$.
\tablefoottext{a}{Akaike  information criterion AIC=$2K+\log L$ \citep{Akaike74}, with $K$ being the number of model parameters.}} 
\label{tab:twocomp}
\end{table}

\subsection{Two-component polarization model} 
\label{sec:twocomponent}

Similarly to \citet{Doroshenko23}, we assume that there are two polarized components in the J0243 data. 
The first   is associated with the pulsar and is described by the RVM. 
The second component is independent of the pulsar phase. 
We express the absolute Stokes parameters for each observation as a sum of the variable and constant components:   
\begin{eqnarray}  
\label{eq:two_comp}
I(\phi) &=& I_{\mathrm c} + I_{\mathrm p}(\phi) , \nonumber \\
Q(\phi) &=& Q_{\mathrm c} + P_{\mathrm p}(\phi)I_{\mathrm p}(\phi)\cos[2\chi(\phi)] , \\
U(\phi) &=& U_{\mathrm c} + P_{\mathrm p}(\phi)I_{\mathrm p}(\phi) \sin[2\chi(\phi)]  .  \nonumber
\end{eqnarray}  
Here we consider the observed Stokes parameters $I$, $Q$, and $U$ to be scaled to the average flux value with indices denoting the constant (c) and pulsed (p) components. 
The PD of the variable component is $P_{\mathrm p}$ and its PA $\chi$ is given by Eq.~\eqref{eq:pa_rvm}.
The scaled Stokes parameters of the constant component $(Q_{\mathrm c},U_{\mathrm c})$ are related to the PD, $P_{\mathrm c}$, and the flux, $I_{\mathrm c}$, as  
\begin{equation}  \label{eq:qu_consta} 
 Q_{\mathrm c}= P_{\mathrm c} I_{\mathrm c} \cos(2\chi_{\mathrm c}),\quad U_{\mathrm c}= P_{\mathrm c} I_{\mathrm c} \sin(2\chi_{\mathrm c}) ,
\end{equation} 
with its PA being $\chi_{\mathrm c}=\frac{1}{2}\arctan(U_{\mathrm c}/Q_{\mathrm c})$.

We assume that the pulsar geometry (i.e., RVM parameters) does not change between the observations and the observed changes in the polarization properties are related to the presence of an additional unpulsed polarized component. 
In order to describe the data from all three observations, in addition to the four RVM parameters, we introduce six additional free parameters $Q_{{\rm c},i}$ and $U_{{\rm c},i}$, $i=1,2,3$, which describe properties of this constant component. 
For a given set of $Q_{{\rm c},i}$ and $U_{{\rm c},i}$, we can construct the differences $Q_{\rm p}=Q(\phi)-Q_{\rm c}$, $U_{\rm p}=U(\phi)-U_{\rm c}$, which are fit by the RVM using maximum likelihood function with the probabilities given by Eq.~\eqref{eq:PA_dist}.
The best-fit RVM parameters as well as the Stokes parameters of the constant component are given in the top part of  Table~\ref{tab:twocomp}. 
The quality of the fit is much better than with the RVM model alone. 
The Akaike information criterion \citep[AIC;][]{Akaike74} decreases by $\Delta\,\mbox{AIC}=21.5$, implying that the pure RVM model is $\exp(\Delta\,\mbox{AIC}/2)=4.7\times10^{4}$ less probable that the two-component model.

\begin{figure}
\centering
\includegraphics[width=0.85\linewidth]{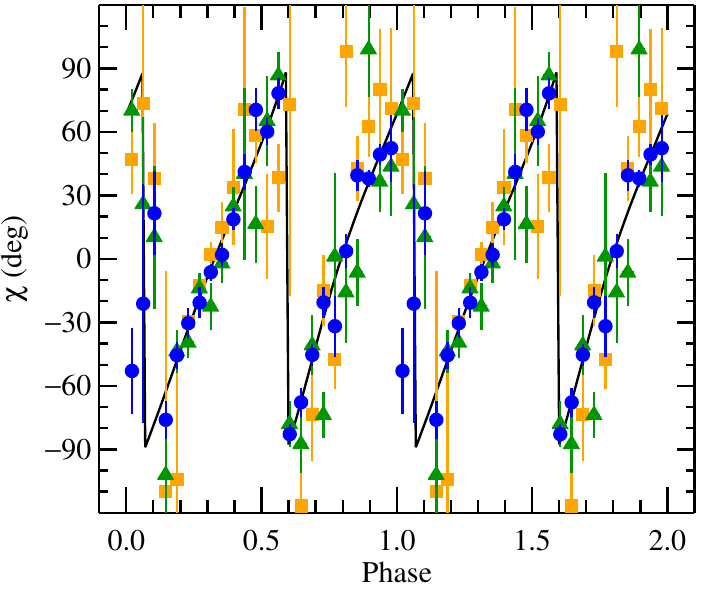}
\caption{Pulse-phase dependence of the PA of the variable (pulsar) components after subtracting the best-fit constant polarized component from the observed Stokes parameters. 
The orange squares, green triangles, and blue circles with error bars correspond to Obs.\,1, 2, and 3, respectively. 
The black line is the best-fit  RVM from the two-component model (see lower part of Table~\ref{tab:twocomp}) to all three data sets. } 
\label{fig:PAintr}
\end{figure}

\begin{figure*}
\centering
\includegraphics[width=0.9\linewidth]{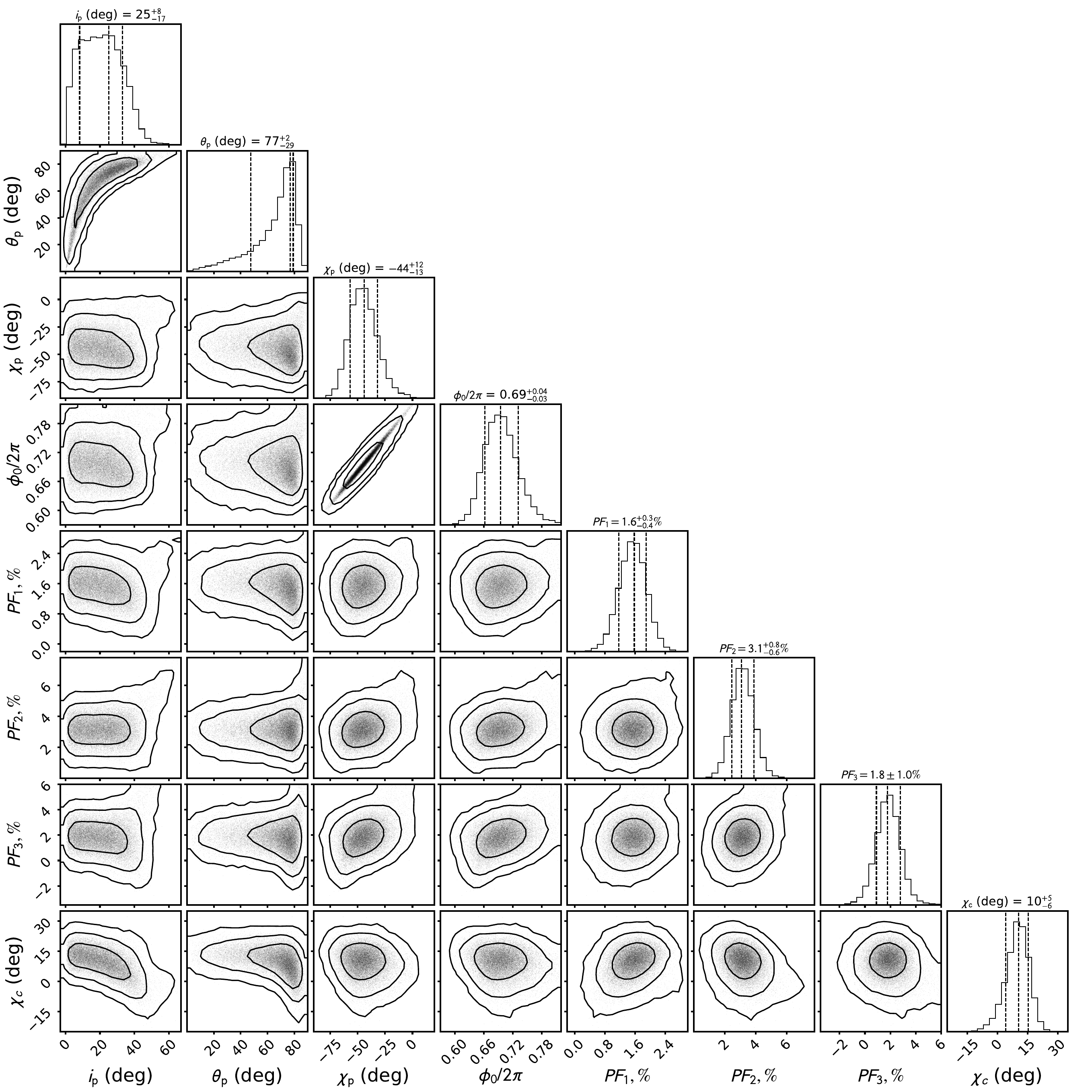}
\caption{Corner plot of the posterior distribution for parameters of the RVM plus a constant model with one free PA given by Eqs.~\eqref{eq:two_comp} fitted directly to the $(q,u)$ values using likelihood function \eqref{eq:PA_dist}. 
Here \textit{PF}$_i,\,i=1$, 2, and 3 is the polarized flux $P_{\rm c}I_{\rm c}$ of the constant component (measured in units of the average flux) during Obs.~1, 2, and 3, respectively, and $\chi_{\rm c}$ is its polarization angle.  
The two-dimensional contours correspond to 68.3\%, 95.45\%, and 99.73\% confidence levels. 
The histograms show the normalized one-dimensional distributions for a given parameter derived from the posterior samples. }
\label{fig:corner_SST}
\end{figure*}

The obtained Stokes parameters of the constant component  correspond to the polarized fluxes (in units of the average flux) of $P_{\mathrm c,i}I_{\mathrm c,i}= \sqrt{Q_{\mathrm c,i}^2+U_{\mathrm c,i}^2}$ 
in the range $\sim$1--4\%. 
Once the best-fit parameters of the constant component are determined, we can obtain the value of the PA for the variable component using Eqs.~\eqref{eq:two_comp}: 
\begin{equation} \label{eq:chi_intr}
\chi(\phi) = \frac{1}{2} \arctan\left[ 
\frac{U(\phi)-U_{\rm c}}{Q(\phi)-Q_{\rm c}}\right] .
\end{equation}
 
We found that the PA values of the constant component $\chi_{\mathrm c,i}$ for the three observations are  the same within the errors. 
This motivated us to perform another fit with the fixed $\chi_{\rm c}$ for all three observations (which  reduced the number of free parameters by 2). 
We refitted the data by varying, in addition to the four RVM parameters, three polarized fluxes $P_{\mathrm c,i}I_{\mathrm c,i}$ and the PA $\chi_{\mathrm c}$. 
The best-fit parameters are presented in the bottom part of  Table~\ref{tab:twocomp}. 
We see that the quality of the fit is not getting much worse,  with $\Delta\,\mbox{AIC}$=2.3, which means that the model is three times less probable. 
The PA of the varying component associated with the pulsar as obtained using Eq.~\eqref{eq:chi_intr} are plotted in Fig.~\ref{fig:PAintr} together with the best-fit RVM.  
In order to obtain the covariance plot for these model parameters we  used the affine invariant Markov chain Monte Carlo ensemble sampler {\sc emcee} package of {\sc python} \citep{emcee13}. 
The resulting posterior distributions are shown in Fig.~\ref{fig:corner_SST}.

\section{Discussion}
\label{sec:disc} 

\subsection{System geometry and origin of constant polarized component}
\label{sec:geometry}

Our fits demonstrated that polarization cannot be described with a simple RVM. 
The main reason is that while the PA  makes two full turns (by 180\degr) during one pulsar period during Obs. 3, it shows a double sine wave during Obs. 1 and 2. 
This  behavior contradicts the RVM. 
Furthermore, the quality of the fit is rather bad and the best-fit pulsar geometrical parameters vary strongly between observations (see Table~\ref{tab:rvm}). 

Following \citet{Doroshenko23},  in addition to the direct pulsar radiation described by the RVM, we assumed that there is a pulse-phase-independent polarized component.  
The data are much better described by this two-component model with a constant set of RVM parameters.   
The PA of the constant component in the three observations was found to lie  between $-20\degr$ and $14\degr$, consistent with  being the same within the errors. 
Fixing this PA at the same value gave the best-fit value of $\chi_{\rm c}= 8\degr\pm7\degr$.  
We found that the polarized flux (in units of the average flux) of the constant component $P_{\rm c}I_{\rm c}$ is between 1.5\% and 3\%. 
Because the contribution of this component to the total flux is unknown, the value of $P_{\rm c}$ is not well determined. 
The lower limit on $P_{\rm c}$ comes from the fact that the flux of the constant component cannot exceed the flux minimum of around 60\% of the average flux (see Fig.~\ref{fig:pulse-profiles}). 
This results in $P_{\rm c}\gtrsim$ 3.5--5\% depending on the observation, with the value growing inversely proportional to $I_{\rm c}$.
A lower limit on $I_{\rm c}$ comes from the condition $P_{\rm c}<100\%$, which translates to $I_{\rm c}\gtrsim$1.5--3\%. 
If we assume more realistically that $P_{\rm c}\lesssim 30\%$  (see below), then $I_{\rm c}\gtrsim$5--10\%.
 
The constant component can be associated with scattering in the equatorial accretion disk wind \citep{Doroshenko23}. 
If the wind half-opening angle is 30\degr, it occupies half of the sky as seen from the central source. 
Thus, for the Thomson optical depth through the wind of 0.2, about 10\% of pulsar radiation is scattered by the wind.  
The PD of the scattered component depends on the disk inclination as $\approx \sin^2i_{\rm d}/(3-\cos^2i_{\rm d})$ (\citealt{ST85}; Nitindala et al. in prep.). 
For $i_{\rm d}>60\degr$, the PD is close to 30\% and the polarized flux in that case has to be around 3\% of the average flux consistent with the data. 
At lower $i_{\rm d}$, a larger contribution of this component to the average flux is required. 

The PA of the constant X-ray component is $\sim$10\degr\ (or $-170\degr$). 
If its polarization is produced in the accretion disk wind, it is natural to assume that the PA is related to the position angle of the accretion disk axis. 
On the other hand, polarization in the optical band is likely produced by scattering of the Be-star radiation off the decretion disk (which occupies a much larger solid angle than the accretion disk around a NS), and therefore provides the orientation of that disk. 
Because of the uncertainty in the value of the IS polarization, the intrinsic optical PA $\chi_{\rm o}$ is in the range between 20\degr\ and 50\degr\ (see Tables~\ref{tab:robopol} and \ref{tab:dipol}). 
The lowest value of $\chi_{\rm o}$ (corresponding to the closest star \#1 as a proxy for IS polarization) is within $2\sigma$ of $\chi_{\rm c}$, while the highest PA is clearly inconsistent with that. 
Thus, there is an indication that the accretion and decretion disks are somewhat misaligned. 

From the two-component model fit, we constrained the pulsar inclination at $i_{\rm p}\approx$ 15\degr--\,40\degr\ and the magnetic obliquity to $\theta_{\rm p}\approx$ 60\degr--\,80\degr. 
The pulsar rotation axis position angle is $\chi_{\rm p}\approx-50\degr\pm10\degr$, which is clearly inconsistent with the PA of the constant component $\chi_{\rm c}$. 
If pulsar radiation escapes in the X-mode, then $\chi_{\rm p}\approx 40\degr$ (or $-140\degr$), which is still far from $\chi_{\rm c}$.  
Under the assumption that $\chi_{\rm c}$ gives the orientation of the accretion disk (and the binary orbit), the difference $|\chi_{\rm p}-\chi_{\rm c}|$ is related to the misalignment angle between the pulsar spin axis and the orbital axis, which is then about 30\degr--\,50\degr. 

We note that evidence for a misalignment was also found   in other pulsars. 
For example, in \mbox{LS V +44 17} \citep{Doroshenko23} we  found  $\chi_{\rm o}\approx \chi_{\rm c}\sim60\degr-\,70\degr$, and both were inconsistent with $\chi_{\rm p}\approx-10\degr$ implying a misalignment of $\sim$75\degr\ if the pulsar emits predominantly in the O-mode or at least 15\degr\ for emission in the X-mode.
In \mbox{Her X-1} \citep{Doroshenko22,Heyl24}, the misalignment is at least 25\degr, but can be $\sim$65\degr\ or even higher.   

Finally, we note that the quality of the fit with the two-component model is still not statistically acceptable, likely indicating that the actual physical situation is even more complex. 
The polarization properties of the additional component may vary with the pulsar phase somehow, but not with the amplitude comparable to that of the pulsar itself. 
For example, scattering in the wind can be phase-dependent because of the azimuthal asymmetry of the pulsar radiation.

\subsection{Polarization--flux anti-correlation}
\label{sec:pdflux}

Our fits demonstrate that most of the polarized flux is produced by the pulsating component. 
Furthermore, the observed polarization has a clear trend of increasing PD with decreasing flux. 
We wondered what the most probable reason for such a behavior could be. 

At low luminosities when most of the radiation is produced in two small hotspots  at the NS surface close to the magnetic poles, the PD is determined by the structure of the atmosphere and the energy dissipation profile \citep{Doroshenko22}.  
The PD is conserved when photons propagate through the magnetosphere and the polarization vector adjusts to the orientation of the magnetic field until the adiabatic radius is reached \citep[e.g.,][]{Heyl02PRD,Taverna15}. 
The adiabatic radius as a function of photon energy $E$ is \citep{Heyl18,Taverna24} 
\begin{equation}
R_{\rm ad}\sim 3\times 10^7\,\left(\frac{E}{1\,{\rm keV}}\right)^{1/5}\,\left(B_{13}R_6^3\right)^{2/5}\,\mbox{cm},  
\end{equation} 
where $B_{13}$ is the surface magnetic field strength in units of $10^{13}$\,G, and $R_6$ is the NS radius in units of $10^6$\,cm. 
For the surface magnetic field of $B_{13}\approx1$ \citep{Kong2022} and $R_6\approx1.2$ \citep[e.g.,][]{Nattila17,Miller20,Annala22}, it exceeds the NS radius for photons in the IXPE range by more than an order of magnitude. 
At this distance, the magnetic field is mostly dipole resulting in the PA that follows the RVM, as indeed observed in a number of X-ray pulsars \citep{Doroshenko22,Malacaria23,Tsygankov2023,Suleimanov2023,Mushtukov23,Heyl24}. 

At a luminosity close to  $10^{38}$\,erg\,s$^{-1}$, a number of additional effects come into play. 
First, the emission region is not point-like anymore, but radiation is produced in an  accretion column, which stands above the NS surface. 
Locally, the polarization direction correlates with the magnetic field direction. 
However, the PD of the whole column radiation can be smaller that the local values, because of the different energy dissipation profile, varying depth of the vacuum resonance \citep{Gnedin78}, and resulting mode conversions \citep{Pavlov79,Lai03ApJ,Doroshenko22,Lai23}. 
Second, a substantial part of the column radiation illuminates the NS surface and is reflected from that \citep{Lyubarskii88,Poutanen13}.
Because of the different relative contributions of the O- and X-modes in the reflected radiation, the total PD is reduced.
Finally, the  accretion flow is expected to be optically thick \citep{Mushtukov17,Mushtukov19}  within the NS magnetospheric radius   
\begin{equation}
R_{\rm m}\simeq 3.4\times 10^8\,\Lambda\,B_{13}^{4/7}\dot{M}_{18}^{-2/7}m^{-1/7}R_6^{12/7}
\approx 2.2\times 10^8\,\mbox{cm},   
\end{equation}
where $\Lambda$ is coefficient typically taken to be $\sim 0.5$ for the case of disk accretion \citep{Chashkina19}, $\dot{M}_{18}$ is the mass accretion rate in units of $10^{18}$\,g\,s$^{-1}$, and $m\approx1.5$ is the mass of a NS in solar masses.
Under this condition, a substantial  fraction of X-ray photons emitted close the NS surface is reprocessed by the optically thick envelope created around the NS magnetospheric cavities.  
Because the size of the magnetosphere exceeds the adiabatic radius, the final polarization of X-ray photons is defined by scattering off the envelope. 
The PD and PA   depend more on the scattering geometry   than on the projection of the NS magnetic dipole on the sky. 
Integrating the Stokes parameters over the envelope  significantly reduces the total PD. 
Thus, in this model the observed anticorrelation of the PD and flux is a result of variable optical thickness of the envelope: the higher the mass accretion rate and the flux, the larger the optical thickness of envelope, and thus the larger the fraction of photons experiencing reprocessing by the envelope outside the NS adiabatic radius that reduces the PD.

\section{Summary}
\label{sec:summary} 

Swift~J0243.6+6124 was observed by IXPE in 2023 July--August  three times during its outburst.  
The main results of our study of its polarimetric properties can be summarized as follows:
\begin{enumerate} 
\item 
Using updated pulsar ephemeris from \textit{Fermi}/GBM we were able to phase connect the pulse arrival times for the whole duration of the outburst. 
\item 
The phase-resolved polarimetric analysis revealed a significant detection of X-ray polarization with the PD reaching $\sim$6\%, 10\%, and 20\% during the three observations separated by a month when the flux dropped by a factor of three.  
\item 
We showed that evolution of the PA with pulsar phase in Obs.\,1 and 2 having a double sine wave structure is inconsistent with the RVM. 
This brought us to the conclusion that the likely reason for this discrepancy is the presence of the phase-independent polarized component   produced, for example by scattering in the accretion disk wind, as was proposed for another bright X-ray pulsar \mbox{LS V +44 17} \citep{Doroshenko23,Poutanen2024}. 
\item 
Assuming the same RVM parameters for the three observations, we fitted the data with the two-component model and obtained the  PA of the constant component $\chi_{\rm c}$ between $-20\degr$ and 15\degr. 
Assuming that the PA for all observations is the same, we find $\chi_{\rm c}=8\degr\pm7\degr$. 
Also assuming that the constant component contributes 10\% of the average flux, we find that the PD of that component varies between 15\% and 30\%. 
We estimated the inclination of the NS rotation axis to the line of sight of $i_{\rm p}=$  15\degr--\,40\degr, the magnetic obliquity of $\theta_{\rm p}=$60\degr--\,80\degr, and the pulsar position angle of $\chi_{\rm p}\approx-50\degr$. 
\item 
Using optical polarimetric observations of the source and nearby field stars, we determined that the intrinsic PA $\chi_{\rm o}$   lies between 20\degr\ and 50\degr, depending on the choice of  field star  to estimate the contribution of the interstellar component. 
The lowest value of $\chi_{\rm o}$ is consistent with $\chi_{\rm c}$ within $2\sigma$, while the higher PA are clearly inconsistent with that value. 
Associating optical polarization with scattering in the decretion disk, the data indicate a possible misalignment between  accretion and decretion disks axes. 
\item A deviation of $\chi_{\rm c}$ from the pulsar position angle $\chi_{\rm p}$ implies a  $\gtrsim30\degr$ misalignment between pulsar rotation axis and the orbital (accretion disk) axis. 
\end{enumerate}

\begin{acknowledgements}
The Imaging X-ray Polarimetry Explorer (IXPE) is a joint US and Italian mission.  The US contribution is supported by the National Aeronautics and Space Administration (NASA) and led and managed by its Marshall Space Flight Center (MSFC), with industry partner Ball Aerospace (contract NNM15AA18C).  The Italian contribution is supported by the Italian Space Agency (Agenzia Spaziale Italiana, ASI) through contract ASI-OHBI-2022-13-I.0, agreements ASI-INAF-2022-19-HH.0 and ASI-INFN-2017.13-H0, and its Space Science Data Center (SSDC) with agreements ASI-INAF-2022-14-HH.0 and ASI-INFN 2021-43-HH.0, and by the Istituto Nazionale di Astrofisica (INAF) and the Istituto Nazionale di Fisica Nucleare (INFN) in Italy.  This research used data products provided by the IXPE Team (MSFC, SSDC, INAF, and INFN) and distributed with additional software tools by the High-Energy Astrophysics Science Archive Research Center (HEASARC), at NASA Goddard Space Flight Center (GSFC).

This research has been supported by the Academy of Finland grants 333112, 349144 and 355672 (JP, SST, AV), the German Academic Exchange Service (DAAD) travel grant 57525212 (VD, VFS), UKRI Stephen Hawking fellowship (AAM), Finnish Cultural Foundation grant (VKra), Natural Sciences and Engineering Research Council of Canada (JH), and Deutsche  Forschungsgemeinschaft (DFG) grant WE 1312/59-1 (VFS).  
IL was supported by the NASA Postdoctoral Program at the Marshall Space Flight Center, administered by Oak Ridge Associated Universities under contract with NASA.
Nordita is supported in part by NordForsk.
\end{acknowledgements}


\begin{thebibliography}{}
\providecommand\natexlab[1]{#1}
\providecommand\JournalTitle[1]{#1}

\bibitem[{{Akaike}(1974)}]{Akaike74}
{Akaike}, H. 1974, \JournalTitle{IEEE Transactions on Automatic Control}, 19,
  716

\bibitem[{{Annala} {et~al.}(2022){Annala}, {Gorda}, {Katerini}, {Kurkela},
  {N{\"a}ttil{\"a}}, {Paschalidis}, \& {Vuorinen}}]{Annala22}
{Annala}, E., {Gorda}, T., {Katerini}, E., {et~al.} 2022,
  \href{http://dx.doi.org/10.1103/PhysRevX.12.011058}{\JournalTitle{Physical
  Review X}, 12, 011058}

\bibitem[{{Arnaud}(1996)}]{Arn96}
{Arnaud}, K.~A. 1996, in ASP Conf. Ser., Vol. 101, Astronomical Data Analysis
  Software and Systems V, ed. G.~H. {Jacoby} \& J.~{Barnes} (San Francisco:
  Astron. Soc. Pac.), 17

\bibitem[{{Bachetti} {et~al.}(2014){Bachetti}, {Harrison}, {Walton},
  {Grefenstette}, {Chakrabarty}, {F{\"u}rst}, {Barret}, {Beloborodov}, {Boggs},
  {Christensen}, {Craig}, {Fabian}, {Hailey}, {Hornschemeier}, {Kaspi},
  {Kulkarni}, {Maccarone}, {Miller}, {Rana}, {Stern}, {Tendulkar}, {Tomsick},
  {Webb}, \& {Zhang}}]{Bachetti2014}
{Bachetti}, M., {Harrison}, F.~A., {Walton}, D.~J., {et~al.} 2014,
  \href{http://dx.doi.org/10.1038/nature13791}{\JournalTitle{\nat}, 514, 202}

\bibitem[{{Bailer-Jones} {et~al.}(2021){Bailer-Jones}, {Rybizki}, {Fouesneau},
  {Demleitner}, \& {Andrae}}]{BailerJones2021}
{Bailer-Jones}, C.~A.~L., {Rybizki}, J., {Fouesneau}, M., {Demleitner}, M., \&
  {Andrae}, R. 2021,
  \href{http://dx.doi.org/10.3847/1538-3881/abd806}{\JournalTitle{\aj}, 161,
  147}

\bibitem[{{Bailer-Jones} {et~al.}(2018){Bailer-Jones}, {Rybizki}, {Fouesneau},
  {Mantelet}, \& {Andrae}}]{BJ18}
{Bailer-Jones}, C.~A.~L., {Rybizki}, J., {Fouesneau}, M., {Mantelet}, G., \&
  {Andrae}, R. 2018,
  \href{http://dx.doi.org/10.3847/1538-3881/aacb21}{\JournalTitle{\aj}, 156,
  58}

\bibitem[{{Baldini} {et~al.}(2021){Baldini}, {Barbanera}, {Bellazzini},
  {Bonino}, {Borotto}, {Brez}, {Caporale}, {Cardelli}, {Castellano},
  {Ceccanti}, {Citraro}, {Di Lalla}, {Latronico}, {Lucchesi}, {Magazz{\`u}},
  {Magazz{\`u}}, {Maldera}, {Manfreda}, {Marengo}, {Marrocchesi}, {Mereu},
  {Minuti}, {Mosti}, {Nasimi}, {Nuti}, {Oppedisano}, {Orsini}, {Pesce-Rollins},
  {Pinchera}, {Profeti}, {Sgr{\`o}}, {Spandre}, {Tardiola}, {Zanetti}, {Amici},
  {Andersson}, {Attin{\`a}}, {Bachetti}, {Baumgartner}, {Brienza},
  {Carpentiero}, {Castronuovo}, {Cavalli}, {Cavazzuti}, {Centrone}, {Costa},
  {D'Alba}, {D'Amico}, {Del Monte}, {Di Cosimo}, {Di Marco}, {Di Persio},
  {Donnarumma}, {Evangelista}, {Fabiani}, {Ferrazzoli}, {Kitaguchi}, {La
  Monaca}, {Lefevre}, {Loffredo}, {Lorenzi}, {Mangraviti}, {Matt}, {Meilahti},
  {Morbidini}, {Muleri}, {Nakano}, {Negri}, {Nenonen}, {O'Dell}, {Perri},
  {Piazzolla}, {Pieraccini}, {Pilia}, {Puccetti}, {Ramsey}, {Rankin},
  {Ratheesh}, {Rubini}, {Santoli}, {Sarra}, {Scalise}, {Sciortino}, {Soffitta},
  {Tamagawa}, {Tennant}, {Tobia}, {Trois}, {Uchiyama}, {Vimercati},
  {Weisskopf}, {Xie}, {Zanetti}, \& {Zhou}}]{2021APh...13302628B}
{Baldini}, L., {Barbanera}, M., {Bellazzini}, R., {et~al.} 2021,
  \href{http://dx.doi.org/10.1016/j.astropartphys.2021.102628}{\JournalTitle{Astroparticle
  Physics}, 133, 102628}

\bibitem[{{Baldini} {et~al.}(2022){Baldini}, {Bucciantini}, {Lalla}, {Ehlert},
  {Manfreda}, {Negro}, {Omodei}, {Pesce-Rollins}, {Sgr{\`o}}, \&
  {Silvestri}}]{Baldini22}
{Baldini}, L., {Bucciantini}, N., {Lalla}, N.~D., {et~al.} 2022,
  \href{http://dx.doi.org/10.1016/j.softx.2022.101194}{\JournalTitle{SoftwareX},
  19, 101194}

\bibitem[{{Basko} \& {Sunyaev}(1976)}]{Basko76}
{Basko}, M.~M., \& {Sunyaev}, R.~A. 1976,
  \href{http://dx.doi.org/10.1093/mnras/175.2.395}{\JournalTitle{\mnras}, 175,
  395}

\bibitem[{{Bikmaev} {et~al.}(2017){Bikmaev}, {Shimansky}, {Irtuganov},
  {Glushkov}, {Sakhibullin}, {Khamitov}, {Burenin}, {Lutovinov}, {Zaznobin},
  {Pavlinsky}, {Sunyaev}, {Dodonov}, {Afanasiev}, {Kotov}, {Doroshenko}, \&
  {Tsygankov}}]{Bikmaev2017}
{Bikmaev}, I., {Shimansky}, V., {Irtuganov}, E., {et~al.} 2017,
  \JournalTitle{The Astronomer's Telegram}, 10968, 1

\bibitem[{{Blackburn}(1995)}]{Blackburn95}
{Blackburn}, J.~K. 1995, in ASP Conf. Ser., Vol.~77, Astronomical Data Analysis
  Software and Systems IV, ed. R.~A. {Shaw}, H.~E. {Payne}, \& J.~J.~E. {Hayes}
  (San Francisco: Astron. Soc. Pac.), 367

\bibitem[{{Brice} {et~al.}(2021){Brice}, {Zane}, {Turolla}, \& {Wu}}]{Brice21}
{Brice}, N., {Zane}, S., {Turolla}, R., \& {Wu}, K. 2021,
  \href{http://dx.doi.org/10.1093/mnras/stab915}{\JournalTitle{\mnras}, 504,
  701}

\bibitem[{{Bykov} {et~al.}(2022){Bykov}, {Gilfanov}, {Tsygankov}, \&
  {Filippova}}]{Bykov22}
{Bykov}, S.~D., {Gilfanov}, M.~R., {Tsygankov}, S.~S., \& {Filippova}, E.~V.
  2022, \href{http://dx.doi.org/10.1093/mnras/stac2239}{\JournalTitle{\mnras},
  516, 1601}

\bibitem[{{Carpano} {et~al.}(2018){Carpano}, {Haberl}, {Maitra}, \&
  {Vasilopoulos}}]{Carpano2018}
{Carpano}, S., {Haberl}, F., {Maitra}, C., \& {Vasilopoulos}, G. 2018,
  \href{http://dx.doi.org/10.1093/mnrasl/sly030}{\JournalTitle{\mnras}, 476,
  L45}

\bibitem[{{Cenko} {et~al.}(2017){Cenko}, {Barthelmy}, {D'Avanzo}, {Kennea},
  {Lien}, {Marshall}, {Palmer}, {Siegel}, \& {Tohuvavohu}}]{Cenko2017}
{Cenko}, S.~B., {Barthelmy}, S.~D., {D'Avanzo}, P., {et~al.} 2017,
  \JournalTitle{GRB Coordinates Network}, 21960, 1

\bibitem[{{Chashkina} {et~al.}(2019){Chashkina}, {Lipunova}, {Abolmasov}, \&
  {Poutanen}}]{Chashkina19}
{Chashkina}, A., {Lipunova}, G., {Abolmasov}, P., \& {Poutanen}, J. 2019,
  \href{http://dx.doi.org/10.1051/0004-6361/201834414}{\JournalTitle{\aap},
  626, A18}

\bibitem[{{Di Marco} {et~al.}(2022){Di Marco}, {Costa}, {Muleri}, {Soffitta},
  {Fabiani}, {La Monaca}, {Rankin}, {Xie}, {Bachetti}, {Baldini},
  {Baumgartner}, {Bellazzini}, {Brez}, {Castellano}, {Del Monte}, {Di Lalla},
  {Ferrazzoli}, {Latronico}, {Maldera}, {Manfreda}, {O'Dell}, {Perri},
  {Pesce-Rollins}, {Puccetti}, {Ramsey}, {Ratheesh}, {Sgr{\`o}}, {Spandre},
  {Tennant}, {Tobia}, {Trois}, \& {Weisskopf}}]{Di_Marco_2022}
{Di Marco}, A., {Costa}, E., {Muleri}, F., {et~al.} 2022,
  \href{http://dx.doi.org/10.3847/1538-3881/ac51c9}{\JournalTitle{\aj}, 163,
  170}

\bibitem[{{Di Marco} {et~al.}(2023){Di Marco}, {Soffitta}, {Costa},
  {Ferrazzoli}, {La Monaca}, {Rankin}, {Ratheesh}, {Xie}, {Baldini}, {Del
  Monte}, {Ehlert}, {Fabiani}, {Kim}, {Muleri}, {O'Dell}, {Ramsey}, {Rubini},
  {Sgr{\`o}}, {Silvestri}, {Tennant}, \& {Weisskopf}}]{Di_Marco_2023}
{Di Marco}, A., {Soffitta}, P., {Costa}, E., {et~al.} 2023,
  \href{http://dx.doi.org/10.3847/1538-3881/acba0f}{\JournalTitle{\aj}, 165,
  143}

\bibitem[{{Doroshenko} {et~al.}(2018){Doroshenko}, {Tsygankov}, \&
  {Santangelo}}]{Doroshenko2018}
{Doroshenko}, V., {Tsygankov}, S., \& {Santangelo}, A. 2018,
  \href{http://dx.doi.org/10.1051/0004-6361/201732208}{\JournalTitle{\aap},
  613, A19}

\bibitem[{{Doroshenko} {et~al.}(2022){Doroshenko}, {Poutanen}, {Tsygankov},
  {Suleimanov}, {Bachetti}, {Caiazzo}, {Costa}, {Di Marco}, {Heyl}, {La
  Monaca}, {Muleri}, {Mushtukov}, {Pavlov}, {Ramsey}, {Rankin}, {Santangelo},
  {Soffitta}, {Staubert}, {Weisskopf}, {Zane}, {Agudo}, {Antonelli}, {Baldini},
  {Baumgartner}, {Bellazzini}, {Bianchi}, {Bongiorno}, {Bonino}, {Brez},
  {Bucciantini}, {Capitanio}, {Castellano}, {Cavazzuti}, {Ciprini}, {De Rosa},
  {Del Monte}, {Di Gesu}, {Di Lalla}, {Donnarumma}, {Dov{\v{c}}iak}, {Ehlert},
  {Enoto}, {Evangelista}, {Fabiani}, {Ferrazzoli}, {Garcia}, {Gunji},
  {Hayashida}, {Iwakiri}, {Jorstad}, {Karas}, {Kitaguchi}, {Kolodziejczak},
  {Krawczynski}, {Latronico}, {Liodakis}, {Maldera}, {Manfreda}, {Marin},
  {Marinucci}, {Marscher}, {Marshall}, {Matt}, {Mitsuishi}, {Mizuno}, {Ng},
  {O'Dell}, {Omodei}, {Oppedisano}, {Papitto}, {Peirson}, {Perri},
  {Pesce-Rollins}, {Pilia}, {Possenti}, {Puccetti}, {Ratheesh}, {Romani},
  {Sgr{\`o}}, {Slane}, {Spandre}, {Sunyaev}, {Tamagawa}, {Tavecchio},
  {Taverna}, {Tawara}, {Tennant}, {Thomas}, {Tombesi}, {Trois}, {Turolla},
  {Vink}, {Wu}, \& {Xie}}]{Doroshenko22}
{Doroshenko}, V., {Poutanen}, J., {Tsygankov}, S.~S., {et~al.} 2022,
  \href{http://dx.doi.org/10.1038/s41550-022-01799-5}{\JournalTitle{Nature
  Astronomy}, 6, 1433}

\bibitem[{{Doroshenko} {et~al.}(2023){Doroshenko}, {Poutanen}, {Heyl},
  {Tsygankov}, {Caiazzo}, {Turolla}, {Veledina}, {Weisskopf}, {Forsblom},
  {Gonz{\'a}lez-Caniulef}, {Loktev}, {Malacaria}, {Mushtukov}, {Suleimanov},
  {Lutovinov}, {Mereminskiy}, {Molkov}, {Salganik}, {Santangelo}, {Berdyugin},
  {Kravtsov}, {Nitindala}, {Agudo}, {Antonelli}, {Bachetti}, {Baldini},
  {Baumgartner}, {Bellazzini}, {Bianchi}, {Bongiorno}, {Bonino}, {Brez},
  {Bucciantini}, {Capitanio}, {Castellano}, {Cavazzuti}, {Chen}, {Ciprini},
  {Costa}, {De Rosa}, {Del Monte}, {Di Gesu}, {Di Lalla}, {Di Marco},
  {Donnarumma}, {Dov{\v{c}}iak}, {Ehlert}, {Enoto}, {Evangelista}, {Fabiani},
  {Ferrazzoli}, {Garc{\'\i}a}, {Gunji}, {Hayashida}, {Iwakiri}, {Jorstad},
  {Kaaret}, {Karas}, {Kislat}, {Kitaguchi}, {Kolodziejczak}, {Krawczynski}, {La
  Monaca}, {Latronico}, {Liodakis}, {Maldera}, {Manfreda}, {Marin},
  {Marinucci}, {Marscher}, {Marshall}, {Massaro}, {Matt}, {Mitsuishi},
  {Mizuno}, {Muleri}, {Negro}, {Ng}, {O'Dell}, {Omodei}, {Oppedisano},
  {Papitto}, {Pavlov}, {Peirson}, {Perri}, {Pesce-Rollins}, {Petrucci},
  {Pilia}, {Possenti}, {Puccetti}, {Ramsey}, {Rankin}, {Ratheesh}, {Roberts},
  {Romani}, {Sgr{\`o}}, {Slane}, {Soffitta}, {Spandre}, {Swartz}, {Tamagawa},
  {Tavecchio}, {Taverna}, {Tawara}, {Tennant}, {Thomas}, {Tombesi}, {Trois},
  {Vink}, {Wu}, {Xie}, \& {Zane}}]{Doroshenko23}
{Doroshenko}, V., {Poutanen}, J., {Heyl}, J., {et~al.} 2023,
  \href{http://dx.doi.org/10.1051/0004-6361/202347088}{\JournalTitle{\aap},
  677, A57}

\bibitem[{{Foreman-Mackey} {et~al.}(2013){Foreman-Mackey}, {Hogg}, {Lang}, \&
  {Goodman}}]{emcee13}
{Foreman-Mackey}, D., {Hogg}, D.~W., {Lang}, D., \& {Goodman}, J. 2013,
  \href{http://dx.doi.org/10.1086/670067}{\JournalTitle{\pasp}, 125, 306}

\bibitem[{{Forsblom} {et~al.}(2023){Forsblom}, {Poutanen}, {Tsygankov},
  {Bachetti}, {Di Marco}, {Doroshenko}, {Heyl}, {La Monaca}, {Malacaria},
  {Marshall}, {Muleri}, {Mushtukov}, {Pilia}, {Rogantini}, {Suleimanov},
  {Taverna}, {Xie}, {Agudo}, {Antonelli}, {Baldini}, {Baumgartner},
  {Bellazzini}, {Bianchi}, {Bongiorno}, {Bonino}, {Brez}, {Bucciantini},
  {Capitanio}, {Castellano}, {Cavazzuti}, {Chen}, {Ciprini}, {Costa}, {De
  Rosa}, {Del Monte}, {Di Gesu}, {Di Lalla}, {Donnarumma}, {Dov{\v{c}}iak},
  {Ehlert}, {Enoto}, {Evangelista}, {Fabiani}, {Ferrazzoli}, {Garcia}, {Gunji},
  {Hayashida}, {Iwakiri}, {Jorstad}, {Kaaret}, {Karas}, {Kitaguchi},
  {Kolodziejczak}, {Krawczynski}, {Latronico}, {Liodakis}, {Maldera},
  {Manfreda}, {Marin}, {Marinucci}, {Marscher}, {Matt}, {Mitsuishi}, {Mizuno},
  {Negro}, {Ng}, {O'Dell}, {Omodei}, {Oppedisano}, {Papitto}, {Pavlov},
  {Peirson}, {Perri}, {Pesce-Rollins}, {Petrucci}, {Possenti}, {Puccetti},
  {Ramsey}, {Rankin}, {Ratheesh}, {Roberts}, {Romani}, {Sgr{\`o}}, {Slane},
  {Soffitta}, {Spandre}, {Sunyaev}, {Swartz}, {Tamagawa}, {Tavecchio},
  {Tawara}, {Tennant}, {Thomas}, {Tombesi}, {Trois}, {Turolla}, {Vink},
  {Weisskopf}, {Wu}, {Zane}, \& {IXPE Collaboration}}]{Forsblom2023}
{Forsblom}, S.~V., {Poutanen}, J., {Tsygankov}, S.~S., {et~al.} 2023,
  \href{http://dx.doi.org/10.3847/2041-8213/acc391}{\JournalTitle{\apjl}, 947,
  L20}

\bibitem[{{Forsblom} {et~al.}(2024){Forsblom}, {Tsygankov}, {Poutanen},
  {Doroshenko}, {Mushtukov}, {Ng}, {Ravi}, {Marshall}, {Di Marco}, {La Monaca},
  {Malacaria}, {Mastroserio}, {Loktev}, {Possenti}, {Suleimanov}, {Taverna},
  {Agudo}, {Antonelli}, {Bachetti}, {Baldini}, {Baumgartner}, {Bellazzini},
  {Bianchi}, {Bongiorno}, {Bonino}, {Brez}, {Bucciantini}, {Capitanio},
  {Castellano}, {Cavazzuti}, {Chen}, {Ciprini}, {Costa}, {De Rosa}, {Del
  Monte}, {Di Gesu}, {Di Lalla}, {Donnarumma}, {Dovciak}, {Ehlert}, {Enoto},
  {Evangelista}, {Fabiani}, {Ferrazzoli}, {Garcia}, {Gunji}, {Hayashida},
  {Heyl}, {Iwakiri}, {Jorstad}, {Kaaret}, {Karas}, {Kislat}, {Kitaguchi},
  {Kolodziejczak}, {Krawczynski}, {Latronico}, {Liodakis}, {Maldera},
  {Manfreda}, {Marin}, {Marinucci}, {Marscher}, {Massaro}, {Matt}, {Mitsuishi},
  {Mizuno}, {Muleri}, {Negro}, {Ng}, {O'Dell}, {Omodei}, {Oppedisano},
  {Papitto}, {Pavlov}, {Peirson}, {Perri}, {Pesce-Rollins}, {Petrucci},
  {Pilia}, {Puccetti}, {Ramsey}, {Rankin}, {Ratheesh}, {Roberts}, {Romani},
  {Sgro}, {Slane}, {Soffitta}, {Spandre}, {Swartz}, {Tamagawa}, {Tavecchio},
  {Tawara}, {Tennant}, {Thomas}, {Tombesi}, {Trois}, {Turolla}, {Vink},
  {Weisskopf}, {Wu}, {Xie}, \& {Zane}}]{Forsblom2024}
{Forsblom}, S.~V., {Tsygankov}, S.~S., {Poutanen}, J., {et~al.} 2024,
  \href{https://doi.org/10.1051/0004-6361/202450937}{\JournalTitle{\aap, in
  press}, arXiv:2406.08988}

\bibitem[{{F{\"u}rst} {et~al.}(2016){F{\"u}rst}, {Walton}, {Harrison}, {Stern},
  {Barret}, {Brightman}, {Fabian}, {Grefenstette}, {Madsen}, {Middleton},
  {Miller}, {Pottschmidt}, {Ptak}, {Rana}, \& {Webb}}]{Furst2016}
{F{\"u}rst}, F., {Walton}, D.~J., {Harrison}, F.~A., {et~al.} 2016,
  \href{http://dx.doi.org/10.3847/2041-8205/831/2/L14}{\JournalTitle{\apjl},
  831, L14}

\bibitem[{{Gehrels} {et~al.}(2004){Gehrels}, {Chincarini}, {Giommi}, {Mason},
  {Nousek}, {Wells}, {White}, {Barthelmy}, {Burrows}, {Cominsky}, {Hurley},
  {Marshall}, {M{\'e}sz{\'a}ros}, {Roming}, {Angelini}, {Barbier}, {Belloni},
  {Campana}, {Caraveo}, {Chester}, {Citterio}, {Cline}, {Cropper}, {Cummings},
  {Dean}, {Feigelson}, {Fenimore}, {Frail}, {Fruchter}, {Garmire}, {Gendreau},
  {Ghisellini}, {Greiner}, {Hill}, {Hunsberger}, {Krimm}, {Kulkarni}, {Kumar},
  {Lebrun}, {Lloyd-Ronning}, {Markwardt}, {Mattson}, {Mushotzky}, {Norris},
  {Osborne}, {Paczynski}, {Palmer}, {Park}, {Parsons}, {Paul}, {Rees},
  {Reynolds}, {Rhoads}, {Sasseen}, {Schaefer}, {Short}, {Smale}, {Smith},
  {Stella}, {Tagliaferri}, {Takahashi}, {Tashiro}, {Townsley}, {Tueller},
  {Turner}, {Vietri}, {Voges}, {Ward}, {Willingale}, {Zerbi}, \&
  {Zhang}}]{Gehrels04}
{Gehrels}, N., {Chincarini}, G., {Giommi}, P., {et~al.} 2004,
  \href{http://dx.doi.org/10.1086/422091}{\JournalTitle{\apj}, 611, 1005}

\bibitem[{{Gnedin} {et~al.}(1978){Gnedin}, {Pavlov}, \& {Shibanov}}]{Gnedin78}
{Gnedin}, Y.~N., {Pavlov}, G.~G., \& {Shibanov}, Y.~A. 1978,
  \JournalTitle{Soviet Astronomy Letters}, 4, 117

\bibitem[{{Heyl} \& {Caiazzo}(2018)}]{Heyl18}
{Heyl}, J., \& {Caiazzo}, I. 2018,
  \href{http://dx.doi.org/10.3390/galaxies6030076}{\JournalTitle{Galaxies}, 6,
  76}

\bibitem[{{Heyl} {et~al.}(2024){Heyl}, {Doroshenko}, {Gonz{\'a}lez-Caniulef},
  {Caiazzo}, {Poutanen}, {Mushtukov}, {Tsygankov}, {Kirmizibayrak}, {Bachetti},
  {Pavlov}, {Forsblom}, {Malacaria}, {Suleimanov}, {Agudo}, {Antonelli},
  {Baldini}, {Baumgartner}, {Bellazzini}, {Bianchi}, {Bongiorno}, {Bonino},
  {Brez}, {Bucciantini}, {Capitanio}, {Castellano}, {Cavazzuti}, {Chen},
  {Ciprini}, {Costa}, {De Rosa}, {Del Monte}, {Di Gesu}, {Di Lalla}, {Di
  Marco}, {Donnarumma}, {Dov{\v{c}}iak}, {Ehlert}, {Enoto}, {Evangelista},
  {Fabiani}, {Ferrazzoli}, {Garcia}, {Gunji}, {Hayashida}, {Iwakiri},
  {Jorstad}, {Kaaret}, {Karas}, {Kislat}, {Kitaguchi}, {Kolodziejczak},
  {Krawczynski}, {La Monaca}, {Latronico}, {Liodakis}, {Maldera}, {Manfreda},
  {Marin}, {Marinucci}, {Marscher}, {Marshall}, {Massaro}, {Matt}, {Mitsuishi},
  {Mizuno}, {Muleri}, {Negro}, {Ng}, {O'Dell}, {Omodei}, {Oppedisano},
  {Papitto}, {Peirson}, {Perri}, {Pesce-Rollins}, {Petrucci}, {Pilia},
  {Possenti}, {Puccetti}, {Ramsey}, {Rankin}, {Ratheesh}, {Roberts}, {Romani},
  {Sgr{\`o}}, {Slane}, {Soffitta}, {Spandre}, {Swartz}, {Tamagawa},
  {Tavecchio}, {Taverna}, {Tawara}, {Tennant}, {Thomas}, {Tombesi}, {Trois},
  {Turolla}, {Vink}, {Weisskopf}, {Wu}, {Xie}, \& {Zane}}]{Heyl24}
{Heyl}, J., {Doroshenko}, V., {Gonz{\'a}lez-Caniulef}, D., {et~al.} 2024,
  \href{http://dx.doi.org/10.1038/s41550-024-02295-8}{\JournalTitle{Nature
  Astronomy}, 8, 1047}

\bibitem[{{Heyl} \& {Shaviv}(2002)}]{Heyl02PRD}
{Heyl}, J.~S., \& {Shaviv}, N.~J. 2002,
  \href{http://dx.doi.org/10.1103/PhysRevD.66.023002}{\JournalTitle{\prd}, 66,
  023002}

\bibitem[{{Heyl} {et~al.}(2003){Heyl}, {Shaviv}, \&
  {Lloyd}}]{2003MNRAS.342..134H}
{Heyl}, J.~S., {Shaviv}, N.~J., \& {Lloyd}, D. 2003,
  \href{http://dx.doi.org/10.1046/j.1365-8711.2003.06521.x}{\JournalTitle{\mnras},
  342, 134}

\bibitem[{{Israel} {et~al.}(2017){Israel}, {Belfiore}, {Stella}, {Esposito},
  {Casella}, {De Luca}, {Marelli}, {Papitto}, {Perri}, {Puccetti}, {Castillo},
  {Salvetti}, {Tiengo}, {Zampieri}, {D'Agostino}, {Greiner}, {Haberl},
  {Novara}, {Salvaterra}, {Turolla}, {Watson}, {Wilms}, \&
  {Wolter}}]{Israel2017}
{Israel}, G.~L., {Belfiore}, A., {Stella}, L., {et~al.} 2017,
  \href{http://dx.doi.org/10.1126/science.aai8635}{\JournalTitle{Science}, 355,
  817}

\bibitem[{{Kaaret} {et~al.}(2017){Kaaret}, {Feng}, \& {Roberts}}]{Kaaret2017}
{Kaaret}, P., {Feng}, H., \& {Roberts}, T.~P. 2017,
  \href{http://dx.doi.org/10.1146/annurev-astro-091916-055259}{\JournalTitle{\araa},
  55, 303}

\bibitem[{{Kennea} {et~al.}(2023){Kennea}, {Bahramian}, \&
  {Negoro}}]{Kennea2023}
{Kennea}, J.~A., {Bahramian}, A., \& {Negoro}, H. 2023, \JournalTitle{The
  Astronomer's Telegram}, 15984, 1

\bibitem[{{Kennea} {et~al.}(2017){Kennea}, {Lien}, {Krimm}, {Cenko}, \&
  {Siegel}}]{Kennea2017}
{Kennea}, J.~A., {Lien}, A.~Y., {Krimm}, H.~A., {Cenko}, S.~B., \& {Siegel},
  M.~H. 2017, \JournalTitle{The Astronomer's Telegram}, 10809, 1

\bibitem[{{King} {et~al.}(2023){King}, {Lasota}, \& {Middleton}}]{King2023}
{King}, A., {Lasota}, J.-P., \& {Middleton}, M. 2023,
  \href{http://dx.doi.org/10.1016/j.newar.2022.101672}{\JournalTitle{\nar}, 96,
  101672}

\bibitem[{{King} {et~al.}(2001){King}, {Davies}, {Ward}, {Fabbiano}, \&
  {Elvis}}]{King01}
{King}, A.~R., {Davies}, M.~B., {Ward}, M.~J., {Fabbiano}, G., \& {Elvis}, M.
  2001, \href{http://dx.doi.org/10.1086/320343}{\JournalTitle{\apjl}, 552,
  L109}

\bibitem[{{Kislat} {et~al.}(2015){Kislat}, {Clark}, {Beilicke}, \&
  {Krawczynski}}]{2015-Kislat}
{Kislat}, F., {Clark}, B., {Beilicke}, M., \& {Krawczynski}, H. 2015,
  \href{http://dx.doi.org/10.1016/j.astropartphys.2015.02.007}{\JournalTitle{Astroparticle
  Physics}, 68, 45}

\bibitem[{{Kong} {et~al.}(2022){Kong}, {Zhang}, {Zhang}, {Ji}, {Doroshenko},
  {Santangelo}, {Chen}, {Lu}, {Ge}, {Wang}, {Tao}, {Qu}, {Li}, {Liu}, {Liao},
  {Chang}, {Peng}, \& {Shui}}]{Kong2022}
{Kong}, L.-D., {Zhang}, S., {Zhang}, S.-N., {et~al.} 2022,
  \href{http://dx.doi.org/10.3847/2041-8213/ac7711}{\JournalTitle{\apjl}, 933,
  L3}

\bibitem[{{Kosenkov}(2021)}]{Kosenkov2021}
{Kosenkov}, I.~A. 2021, 
  \href{http://dx.doi.org/10.5281/zenodo.5763988}{Dipol2Red: Linear polarization data reduction pipeline
  for DIPol-2 and DIPol-UF polarimeters} 
 


\bibitem[{{Kosenkov} {et~al.}(2017){Kosenkov}, {Berdyugin}, {Piirola},
  {Tsygankov}, {Pall{\'e}}, {Miles-P{\'a}ez}, \& {Poutanen}}]{Kosenkov2017}
{Kosenkov}, I.~A., {Berdyugin}, A.~V., {Piirola}, V., {et~al.} 2017,
  \href{http://dx.doi.org/10.1093/mnras/stx779}{\JournalTitle{\mnras}, 468,
  4362}

\bibitem[{{Kouroubatzakis} {et~al.}(2017){Kouroubatzakis}, {Reig}, {Andrews},
  \& {Zezas}}]{Kouroubatzakis2017}
{Kouroubatzakis}, K., {Reig}, P., {Andrews}, J., \& {Zezas}, A. 2017,
  \JournalTitle{The Astronomer's Telegram}, 10822, 1

\bibitem[{{Lai}(2023)}]{Lai23}
{Lai}, D. 2023,
  \href{http://dx.doi.org/10.1073/pnas.2216534120}{\JournalTitle{PNAS}, 120,
  e2216534120}

\bibitem[{{Lai} \& {Ho}(2003)}]{Lai03ApJ}
{Lai}, D., \& {Ho}, W. C.~G. 2003,
  \href{http://dx.doi.org/10.1086/374334}{\JournalTitle{\apj}, 588, 962}

\bibitem[{{Lipunov}(1982)}]{Lipunov82}
{Lipunov}, V.~M. 1982, \JournalTitle{\sovast}, 26, 54

\bibitem[{{Loktev} {et~al.}(2020){Loktev}, {Salmi}, {N{\"a}ttil{\"a}}, \&
  {Poutanen}}]{Loktev20}
{Loktev}, V., {Salmi}, T., {N{\"a}ttil{\"a}}, J., \& {Poutanen}, J. 2020,
  \href{http://dx.doi.org/10.1051/0004-6361/202039134}{\JournalTitle{\aap},
  643, A84}

\bibitem[{{Lyubarskii} \& {Syunyaev}(1988)}]{Lyubarskii88}
{Lyubarskii}, Y.~E., \& {Syunyaev}, R.~A. 1988, \JournalTitle{Soviet Astronomy
  Letters}, 14, 390

\bibitem[{{Majumder} {et~al.}(2024){Majumder}, {Chatterjee}, {Jayasurya},
  {Das}, \& {Nandi}}]{Majumder24}
{Majumder}, S., {Chatterjee}, R., {Jayasurya}, K.~M., {Das}, S., \& {Nandi}, A.
  2024, \href{http://dx.doi.org/10.3847/2041-8213/ad67e5}{\JournalTitle{\apjl},
  971, L21}

\bibitem[{{Malacaria} {et~al.}(2023){Malacaria}, {Heyl}, {Doroshenko},
  {Tsygankov}, {Poutanen}, {Forsblom}, {Capitanio}, {Di Marco}, {Du}, {Ducci},
  {La Monaca}, {Lutovinov}, {Marshall}, {Mereminskiy}, {Molkov}, {Ng},
  {Petrucci}, {Santangelo}, {Shtykovsky}, {Suleimanov}, {Agudo}, {Antonelli},
  {Bachetti}, {Baldini}, {Baumgartner}, {Bellazzini}, {Bianchi}, {Bongiorno},
  {Bonino}, {Brez}, {Bucciantini}, {Castellano}, {Cavazzuti}, {Chen},
  {Ciprini}, {Costa}, {De Rosa}, {Del Monte}, {Di Gesu}, {Di Lalla},
  {Donnarumma}, {Dovciak}, {Ehlert}, {Enoto}, {Evangelista}, {Fabiani},
  {Ferrazzoli}, {Garcia}, {Gunji}, {Hayashida}, {Iwakiri}, {Jorstad}, {Kaaret},
  {Karas}, {Kislat}, {Kitaguchi}, {Kolodziejczak1}, {Krawczynski}, {Latronico},
  {Liodakis}, {Maldera}, {Manfreda}, {Marin}, {Marinucci}, {Marscher},
  {Massaro}, {Matt}, {Mitsuishi}, {Mizuno}, {Muleri}, {Negro}, {Ng}, {O'Dell},
  {Omodei}, {Oppedisano}, {Papitto}, {Pavlov}, {Peirson}, {Perri},
  {Pesce-Rollins}, {Pilia}, {Possenti}, {Puccetti}, {Ramsey}, {Rankin},
  {Ratheesh}, {Roberts}, {Romani}, {Sgro}, {Slane}, {Soffitta}, {Spandre},
  {Swartz}, {Tamagawa}, {Tavecchio}, {Taverna}, {Tawara}, {Tennant}, {Thomas},
  {Tombesi}, {Trois}, {Turolla}, {Vink}, {Weisskopf}, {Wu}, {Xie}, \&
  {Zane}}]{Malacaria23}
{Malacaria}, C., {Heyl}, J., {Doroshenko}, V., {et~al.} 2023,
  \href{http://dx.doi.org/10.1051/0004-6361/202346581}{\JournalTitle{\aap},
  675, A29}

\bibitem[{{Marin} {et~al.}(2024){Marin}, {Marinucci}, {Laurenti}, {Kim},
  {Barnouin}, {Di Marco}, {Ursini}, {Bianchi}, {Ravi}, {Marshall}, {Matt},
  {Chen}, {Gianolli}, {Ingram}, {Maksym}, {Panagiotou}, {Podgorny}, {Puccetti},
  {Ratheesh}, {Tombesi}, {Agudo}, {Antonelli}, {Bachetti}, {Baldini},
  {Baumgartner}, {Bellazzini}, {Bongiorno}, {Bonino}, {Brez}, {Bucciantini},
  {Capitanio}, {Castellano}, {Cavazzuti}, {Ciprini}, {Costa}, {De Rosa}, {Del
  Monte}, {Di Gesu}, {Di Lalla}, {Donnarumma}, {Doroshenko}, {Dovciak},
  {Ehlert}, {Enoto}, {Evangelista}, {Fabiani}, {Ferrazzoli}, {Garcia}, {Gunji},
  {Heyl}, {Iwakiri}, {Jorstad}, {Kaaret}, {Karas}, {Kislat}, {Kitaguchi},
  {Kolodziejczak}, {Krawczynski}, {La Monaca}, {Latronico}, {Liodakis},
  {Madejski}, {Maldera}, {Manfreda}, {Marscher}, {Massaro}, {Mitsuishi},
  {Mizuno}, {Muleri}, {Negro}, {Ng}, {O'Dell}, {Omodei}, {Oppedisano},
  {Papitto}, {Pavlov}, {Perri}, {Pesce-Rollins}, {Petrucci}, {Pilia},
  {Possenti}, {Poutanen}, {Ramsey}, {Rankin}, {Roberts}, {Romani}, {Sgro},
  {Slane}, {Soffitta}, {Spandre}, {Swartz}, {Tamagawa}, {Tavecchio}, {Taverna},
  {Tawara}, {Tennant}, {Thomas}, {Trois}, {Tsygankov}, {Turolla}, {Vink},
  {Weisskopf}, {Wu}, {Xie}, \& {Zane}}]{Marin24}
{Marin}, F., {Marinucci}, A., {Laurenti}, M., {et~al.} 2024,
  \href{http://dx.doi.org/10.1051/0004-6361/202449760}{\JournalTitle{\aap}, 689, A238}

\bibitem[{{Matsuoka} {et~al.}(2009){Matsuoka}, {Kawasaki}, {Ueno}, {Tomida},
  {Kohama}, {Suzuki}, {Adachi}, {Ishikawa}, {Mihara}, {Sugizaki}, {Isobe},
  {Nakagawa}, {Tsunemi}, {Miyata}, {Kawai}, {Kataoka}, {Morii}, {Yoshida},
  {Negoro}, {Nakajima}, {Ueda}, {Chujo}, {Yamaoka}, {Yamazaki}, {Nakahira},
  {You}, {Ishiwata}, {Miyoshi}, {Eguchi}, {Hiroi}, {Katayama}, \&
  {Ebisawa}}]{Matsuoka2009}
{Matsuoka}, M., {Kawasaki}, K., {Ueno}, S., {et~al.} 2009,
  \href{http://dx.doi.org/10.1093/pasj/61.5.999}{\JournalTitle{\pasj}, 61, 999}

\bibitem[{{Matt}(1993)}]{Matt93}
{Matt}, G. 1993,
  \href{http://dx.doi.org/10.1093/mnras/260.3.663}{\JournalTitle{\mnras}, 260,
  663}

\bibitem[{{Meszaros} {et~al.}(1988){Meszaros}, {Novick}, {Szentgyorgyi},
  {Chanan}, \& {Weisskopf}}]{Meszaros88}
{Meszaros}, P., {Novick}, R., {Szentgyorgyi}, A., {Chanan}, G.~A., \&
  {Weisskopf}, M.~C. 1988,
  \href{http://dx.doi.org/10.1086/165962}{\JournalTitle{\apj}, 324, 1056}

\bibitem[{{Miller} {et~al.}(2020){Miller}, {Chirenti}, \& {Lamb}}]{Miller20}
{Miller}, M.~C., {Chirenti}, C., \& {Lamb}, F.~K. 2020,
  \href{http://dx.doi.org/10.3847/1538-4357/ab4ef9}{\JournalTitle{\apj}, 888,
  12}

\bibitem[{{Mushtukov} {et~al.}(2019){Mushtukov}, {Ingram}, {Middleton},
  {Nagirner}, \& {van der Klis}}]{Mushtukov19}
{Mushtukov}, A.~A., {Ingram}, A., {Middleton}, M., {Nagirner}, D.~I., \& {van
  der Klis}, M. 2019,
  \href{http://dx.doi.org/10.1093/mnras/sty3525}{\JournalTitle{\mnras}, 484,
  687}

\bibitem[{{Mushtukov} \& {Portegies Zwart}(2023)}]{Mushtukov23b}
{Mushtukov}, A.~A., \& {Portegies Zwart}, S. 2023,
  \href{http://dx.doi.org/10.1093/mnras/stac3431}{\JournalTitle{\mnras}, 518,
  5457}

\bibitem[{{Mushtukov} {et~al.}(2021){Mushtukov}, {Portegies Zwart},
  {Tsygankov}, {Nagirner}, \& {Poutanen}}]{Mushtukov21}
{Mushtukov}, A.~A., {Portegies Zwart}, S., {Tsygankov}, S.~S., {Nagirner},
  D.~I., \& {Poutanen}, J. 2021,
  \href{http://dx.doi.org/10.1093/mnras/staa3809}{\JournalTitle{\mnras}, 501,
  2424}

\bibitem[{{Mushtukov} {et~al.}(2017){Mushtukov}, {Suleimanov}, {Tsygankov}, \&
  {Ingram}}]{Mushtukov17}
{Mushtukov}, A.~A., {Suleimanov}, V.~F., {Tsygankov}, S.~S., \& {Ingram}, A.
  2017, \href{http://dx.doi.org/10.1093/mnras/stx141}{\JournalTitle{\mnras},
  467, 1202}

\bibitem[{{Mushtukov} {et~al.}(2015){Mushtukov}, {Suleimanov}, {Tsygankov}, \&
  {Poutanen}}]{Mushtukov15}
{Mushtukov}, A.~A., {Suleimanov}, V.~F., {Tsygankov}, S.~S., \& {Poutanen}, J.
  2015, \href{http://dx.doi.org/10.1093/mnras/stu2484}{\JournalTitle{\mnras},
  447, 1847}

\bibitem[{{Mushtukov} {et~al.}(2023){Mushtukov}, {Tsygankov}, {Poutanen},
  {Doroshenko}, {Salganik}, {Costa}, {Di Marco}, {Heyl}, {La Monaca},
  {Lutovinov}, {Mereminsky}, {Papitto}, {Semena}, {Shtykovsky}, {Suleimanov},
  {Forsblom}, {Gonz{\'a}lez-Caniulef}, {Malacaria}, {Sunyaev}, {Agudo},
  {Antonelli}, {Bachetti}, {Baldini}, {Baumgartner}, {Bellazzini}, {Bianchi},
  {Bongiorno}, {Bonino}, {Brez}, {Bucciantini}, {Capitanio}, {Castellano},
  {Cavazzuti}, {Chen}, {Ciprini}, {De Rosa}, {Del Monte}, {Di Gesu}, {Di
  Lalla}, {Donnarumma}, {Dov{\v{c}}iak}, {Ehlert}, {Enoto}, {Evangelista},
  {Fabiani}, {Ferrazzoli}, {Garcia}, {Gunji}, {Hayashida}, {Iwakiri},
  {Jorstad}, {Kaaret}, {Karas}, {Kislat}, {Kitaguchi}, {Kolodziejczak},
  {Krawczynski}, {Latronico}, {Liodakis}, {Maldera}, {Manfreda}, {Marin},
  {Marscher}, {Marshall}, {Massaro}, {Matt}, {Mitsuishi}, {Mizuno}, {Muleri},
  {Negro}, {Ng}, {O'Dell}, {Omodei}, {Oppedisano}, {Pavlov}, {Peirson},
  {Perri}, {Pesce-Rollins}, {Petrucci}, {Pilia}, {Possenti}, {Puccetti},
  {Ramsey}, {Rankin}, {Ratheesh}, {Roberts}, {Romani}, {Sgr{\`o}}, {Slane},
  {Soffitta}, {Spandre}, {Swartz}, {Tamagawa}, {Tavecchio}, {Taverna},
  {Tawara}, {Tennant}, {Thomas}, {Tombesi}, {Trois}, {Turolla}, {Vink},
  {Weisskopf}, {Wu}, {Xie}, \& {Zane}}]{Mushtukov23}
{Mushtukov}, A.~A., {Tsygankov}, S.~S., {Poutanen}, J., {et~al.} 2023,
  \href{http://dx.doi.org/10.1093/mnras/stad1961}{\JournalTitle{\mnras}, 524,
  2004}

\bibitem[{{Naghizadeh-Khouei} \& {Clarke}(1993)}]{Naghizadeh1993}
{Naghizadeh-Khouei}, J., \& {Clarke}, D. 1993, \JournalTitle{\aap}, 274, 968

\bibitem[{{N{\"a}ttil{\"a}} {et~al.}(2017){N{\"a}ttil{\"a}}, {Miller},
  {Steiner}, {Kajava}, {Suleimanov}, \& {Poutanen}}]{Nattila17}
{N{\"a}ttil{\"a}}, J., {Miller}, M.~C., {Steiner}, A.~W., {et~al.} 2017,
  \href{http://dx.doi.org/10.1051/0004-6361/201731082}{\JournalTitle{\aap},
  608, A31}

\bibitem[{{Ng} {et~al.}(2023){Ng}, {Sanna}, {Chakrabarty}, {Malacaria},
  {Jaisawal}, {Pradhan}, {Coley}, {Wolff}, {Guillot}, {Arzoumanian},
  {Gendreau}, \& {Ferrara}}]{Ng2023}
{Ng}, M., {Sanna}, A., {Chakrabarty}, D., {et~al.} 2023, \JournalTitle{The
  Astronomer's Telegram}, 15987, 1

\bibitem[{{Pavlov} \& {Shibanov}(1979)}]{Pavlov79}
{Pavlov}, G.~G., \& {Shibanov}, Y.~A. 1979, \JournalTitle{JETP}, 49, 741

\bibitem[{{Piirola} {et~al.}(2014){Piirola}, {Berdyugin}, \&
  {Berdyugina}}]{Piirola2014}
{Piirola}, V., {Berdyugin}, A., \& {Berdyugina}, S. 2014,
  \href{http://dx.doi.org/10.1117/12.2055923}{in \procspie, Vol. 9147,
  Ground-based and Airborne Instrumentation for Astronomy V}, 91478I

\bibitem[{{Poutanen}(2020)}]{Poutanen2020}
{Poutanen}, J. 2020,
  \href{http://dx.doi.org/10.1051/0004-6361/202038689}{\JournalTitle{\aap},
  641, A166}

\bibitem[{{Poutanen} {et~al.}(2007){Poutanen}, {Lipunova}, {Fabrika},
  {Butkevich}, \& {Abolmasov}}]{Poutanen07}
{Poutanen}, J., {Lipunova}, G., {Fabrika}, S., {Butkevich}, A.~G., \&
  {Abolmasov}, P. 2007,
  \href{http://dx.doi.org/10.1111/j.1365-2966.2007.11668.x}{\JournalTitle{\mnras},
  377, 1187}

\bibitem[{{Poutanen} {et~al.}(2013){Poutanen}, {Mushtukov}, {Suleimanov},
  {Tsygankov}, {Nagirner}, {Doroshenko}, \& {Lutovinov}}]{Poutanen13}
{Poutanen}, J., {Mushtukov}, A.~A., {Suleimanov}, V.~F., {et~al.} 2013,
  \href{http://dx.doi.org/10.1088/0004-637X/777/2/115}{\JournalTitle{\apj},
  777, 115}

\bibitem[{{Poutanen} {et~al.}(1996){Poutanen}, {Nagendra}, \&
  {Svensson}}]{PNS96}
{Poutanen}, J., {Nagendra}, K.~N., \& {Svensson}, R. 1996,
  \href{http://dx.doi.org/10.1093/mnras/283.3.892}{\JournalTitle{\mnras}, 283,
  892}

\bibitem[{{Poutanen} {et~al.}(2024){Poutanen}, {Tsygankov}, \&
  {Forsblom}}]{Poutanen2024}
{Poutanen}, J., {Tsygankov}, S.~S., \& {Forsblom}, S.~V. 2024,
  \href{http://dx.doi.org/10.3390/galaxies12040046}{\JournalTitle{Galaxies},
  12, 46}

\bibitem[{{Radhakrishnan} \& {Cooke}(1969)}]{Radhakrishnan69}
{Radhakrishnan}, V., \& {Cooke}, D.~J. 1969, \JournalTitle{\aplett}, 3, 225

\bibitem[{{Rodr{\'\i}guez Castillo}(2020)}]{RodriguezCastillo2020}
{Rodr{\'\i}guez Castillo}, G.~A. 2020,
  \href{http://dx.doi.org/10.3847/1538-4357/ab8a44}{\JournalTitle{\apj}, 895,
  60}

\bibitem[{{Salganik} {et~al.}(2023){Salganik}, {Tsygankov}, {Doroshenko},
  {Molkov}, {Lutovinov}, {Mushtukov}, \& {Poutanen}}]{Salganik23}
{Salganik}, A., {Tsygankov}, S.~S., {Doroshenko}, V., {et~al.} 2023,
  \href{http://dx.doi.org/10.1093/mnras/stad2124}{\JournalTitle{\mnras}, 524,
  5213}

\bibitem[{{Setoguchi} {et~al.}(2023){Setoguchi}, {Negoro}, {Nakajima},
  {Kobayashi}, {Tanaka}, {Soejima}, {Mihara}, {Kawamuro}, {Yamada}, {Tamagawa},
  {Matsuoka}, {Sakamoto}, {Serino}, {Sugita}, {Hiramatsu}, {Nishikawa},
  {Yoshida}, {Tsuboi}, {Urabe}, {Nawa}, {Nemoto}, {Shidatsu}, {Iwasaki},
  {Kawai}, {Niwano}, {Hosokawa}, {Imai}, {Ito}, {Takamatsu}, {Nakahira},
  {Ueno}, {Tomida}, {Ishikawa}, {Kurihara}, {Ueda}, {Ogawa}, {Yoshitake},
  {Inaba}, {Nakatani}, {Yamauchi}, {Sato}, {Hatsuda}, {Fukuoka}, {Hagiwara},
  {Umeki}, {Otsuki}, {Yamaoka}, {Kawakubo}, {Sugizaki}, \&
  {Iwakiri}}]{Setoguchi2023}
{Setoguchi}, K., {Negoro}, H., {Nakajima}, M., {et~al.} 2023, \JournalTitle{The
  Astronomer's Telegram}, 15983, 1

\bibitem[{{Soffitta} {et~al.}(2021){Soffitta}, {Baldini}, {Bellazzini},
  {Costa}, {Latronico}, {Muleri}, {Del Monte}, {Fabiani}, {Minuti}, {Pinchera},
  {Sgro'}, {Spandre}, {Trois}, {Amici}, {Andersson}, {Attina'}, {Bachetti},
  {Barbanera}, {Borotto}, {Brez}, {Brienza}, {Caporale}, {Cardelli},
  {Carpentiero}, {Castellano}, {Castronuovo}, {Cavalli}, {Cavazzuti},
  {Ceccanti}, {Centrone}, {Ciprini}, {Citraro}, {D'Amico}, {D'Alba}, {Di
  Cosimo}, {Di Lalla}, {Di Marco}, {Di Persio}, {Donnarumma}, {Evangelista},
  {Ferrazzoli}, {Hayato}, {Kitaguchi}, {La Monaca}, {Lefevre}, {Loffredo},
  {Lorenzi}, {Lucchesi}, {Magazzu}, {Maldera}, {Manfreda}, {Mangraviti},
  {Marengo}, {Matt}, {Mereu}, {Morbidini}, {Mosti}, {Nakano}, {Nasimi},
  {Negri}, {Nenonen}, {Nuti}, {Orsini}, {Perri}, {Pesce-Rollins}, {Piazzolla},
  {Pilia}, {Profeti}, {Puccetti}, {Rankin}, {Ratheesh}, {Rubini}, {Santoli},
  {Sarra}, {Scalise}, {Sciortino}, {Tamagawa}, {Tardiola}, {Tobia},
  {Vimercati}, \& {Xie}}]{2021AJ....162..208S}
{Soffitta}, P., {Baldini}, L., {Bellazzini}, R., {et~al.} 2021,
  \href{http://dx.doi.org/10.3847/1538-3881/ac19b0}{\JournalTitle{\aj}, 162,
  208}

\bibitem[{{Suleimanov} {et~al.}(2023){Suleimanov}, {Forsblom}, {Tsygankov},
  {Poutanen}, {Doroshenko}, {Doroshenko}, {Capitanio}, {Di Marco},
  {Gonz{\'a}lez-Caniulef}, {Heyl}, {La Monaca}, {Lutovinov}, {Molkov},
  {Malacaria}, {Mushtukov}, {Shtykovsky}, {Agudo}, {Antonelli}, {Bachetti},
  {Baldini}, {Baumgartner}, {Bellazzini}, {Bianchi}, {Bongiorno}, {Bonino},
  {Brez}, {Bucciantini}, {Castellano}, {Cavazzuti}, {Chen}, {Ciprini}, {Costa},
  {De Rosa}, {Del Monte}, {Di Gesu}, {Di Lalla}, {Donnarumma}, {Dov{\v{c}}iak},
  {Ehlert}, {Enoto}, {Evangelista}, {Fabiani}, {Ferrazzoli}, {Garcia}, {Gunji},
  {Hayashida}, {Iwakiri}, {Jorstad}, {Kaaret}, {Karas}, {Kislat}, {Kitaguchi},
  {Kolodziejczak}, {Krawczynski}, {Latronico}, {Liodakis}, {Maldera},
  {Manfreda}, {Marin}, {Marinucci}, {Marscher}, {Marshall}, {Massaro}, {Matt},
  {Mitsuishi}, {Mizuno}, {Muleri}, {Negro}, {Ng}, {O'Dell}, {Omodei},
  {Oppedisano}, {Papitto}, {Pavlov}, {Peirson}, {Perri}, {Pesce-Rollins},
  {Petrucci}, {Pilia}, {Possenti}, {Puccetti}, {Ramsey}, {Rankin}, {Ratheesh},
  {Roberts}, {Romani}, {Sgr{\`o}}, {Slane}, {Soffitta}, {Spandre}, {Swartz},
  {Tamagawa}, {Tavecchio}, {Taverna}, {Tawara}, {Tennant}, {Thomas}, {Tombesi},
  {Trois}, {Turolla}, {Vink}, {Weisskopf}, {Wu}, {Xie}, \&
  {Zane}}]{Suleimanov2023}
{Suleimanov}, V.~F., {Forsblom}, S.~V., {Tsygankov}, S.~S., {et~al.} 2023,
  \href{http://dx.doi.org/10.1051/0004-6361/202346994}{\JournalTitle{\aap},
  678, A119}

\bibitem[{{Sunyaev} \& {Titarchuk}(1985)}]{ST85}
{Sunyaev}, R.~A., \& {Titarchuk}, L.~G. 1985,   \href{https://ui.adsabs.harvard.edu/abs/1985A%26A...143..374S/abstract}{\JournalTitle{\aap}, 143, 374}


\bibitem[{{Taverna} \& {Turolla}(2024)}]{Taverna24}
{Taverna}, R., \& {Turolla}, R. 2024,
  \href{http://dx.doi.org/10.3390/galaxies12010006}{\JournalTitle{Galaxies},
  12, 6}

\bibitem[{{Taverna} {et~al.}(2015){Taverna}, {Turolla}, {Gonzalez Caniulef},
  {Zane}, {Muleri}, \& {Soffitta}}]{Taverna15}
{Taverna}, R., {Turolla}, R., {Gonzalez Caniulef}, D., {et~al.} 2015,
  \href{http://dx.doi.org/10.1093/mnras/stv2168}{\JournalTitle{\mnras}, 454,
  3254}

\bibitem[{{Tsygankov} {et~al.}(2018){Tsygankov}, {Doroshenko}, {Mushtukov},
  {Lutovinov}, \& {Poutanen}}]{Tsygankov2018}
{Tsygankov}, S.~S., {Doroshenko}, V., {Mushtukov}, A.~A., {Lutovinov}, A.~A.,
  \& {Poutanen}, J. 2018,
  \href{http://dx.doi.org/10.1093/mnrasl/sly116}{\JournalTitle{\mnras}, 479,
  L134}

\bibitem[{{Tsygankov} {et~al.}(2022){Tsygankov}, {Doroshenko}, {Poutanen},
  {Heyl}, {Mushtukov}, {Caiazzo}, {Di Marco}, {Forsblom},
  {Gonz{\'a}lez-Caniulef}, {Klawin}, {La Monaca}, {Malacaria}, {Marshall},
  {Muleri}, {Ng}, {Suleimanov}, {Sunyaev}, {Turolla}, {Agudo}, {Antonelli},
  {Bachetti}, {Baldini}, {Baumgartner}, {Bellazzini}, {Bianchi}, {Bongiorno},
  {Bonino}, {Brez}, {Bucciantini}, {Capitanio}, {Castellano}, {Cavazzuti},
  {Ciprini}, {Costa}, {De Rosa}, {Del Monte}, {Di Gesu}, {Di Lalla},
  {Donnarumma}, {Dov{\v{c}}iak}, {Ehlert}, {Enoto}, {Evangelista}, {Fabiani},
  {Ferrazzoli}, {Garcia}, {Gunji}, {Hayashida}, {Iwakiri}, {Jorstad}, {Karas},
  {Kitaguchi}, {Kolodziejczak}, {Krawczynski}, {Latronico}, {Liodakis},
  {Maldera}, {Manfreda}, {Marin}, {Marinucci}, {Marscher}, {Matt}, {Mitsuishi},
  {Mizuno}, {Ng}, {O'Dell}, {Omodei}, {Oppedisano}, {Papitto}, {Pavlov},
  {Peirson}, {Perri}, {Pesce-Rollins}, {Petrucci}, {Pilia}, {Possenti},
  {Puccetti}, {Ramsey}, {Rankin}, {Ratheesh}, {Romani}, {Sgr{\`o}}, {Slane},
  {Soffitta}, {Spandre}, {Tamagawa}, {Tavecchio}, {Taverna}, {Tawara},
  {Tennant}, {Thomas}, {Tombesi}, {Trois}, {Vink}, {Weisskopf}, {Wu}, {Xie},
  {Zane}, \& {IXPE Collaboration}}]{Tsygankov22}
{Tsygankov}, S.~S., {Doroshenko}, V., {Poutanen}, J., {et~al.} 2022,
  \href{http://dx.doi.org/10.3847/2041-8213/aca486}{\JournalTitle{\apjl}, 941,
  L14}

\bibitem[{{Tsygankov} {et~al.}(2023){Tsygankov}, {Doroshenko}, {Mushtukov},
  {Poutanen}, {Di Marco}, {Heyl}, {La Monaca}, {Forsblom}, {Malacaria},
  {Marshall}, {Suleimanov}, {Svoboda}, {Taverna}, {Ursini}, {Agudo},
  {Antonelli}, {Bachetti}, {Baldini}, {Baumgartner}, {Bellazzini}, {Bianchi},
  {Bongiorno}, {Bonino}, {Brez}, {Bucciantini}, {Capitanio}, {Castellano},
  {Cavazzuti}, {Chen}, {Ciprini}, {Costa}, {De Rosa}, {Del Monte}, {Di Gesu},
  {Di Lalla}, {Donnarumma}, {Dov{\v{c}}iak}, {Ehlert}, {Enoto}, {Evangelista},
  {Fabiani}, {Ferrazzoli}, {Garcia}, {Gunji}, {Hayashida}, {Iwakiri},
  {Jorstad}, {Kaaret}, {Karas}, {Kislat}, {Kitaguchi}, {Kolodziejczak},
  {Krawczynski}, {Latronico}, {Liodakis}, {Maldera}, {Manfreda}, {Marin},
  {Marinucci}, {Marscher}, {Massaro}, {Matt}, {Mitsuishi}, {Mizuno}, {Muleri},
  {Negro}, {Ng}, {O'Dell}, {Omodei}, {Oppedisano}, {Papitto}, {Pavlov},
  {Peirson}, {Perri}, {Pesce-Rollins}, {Petrucci}, {Pilia}, {Possenti},
  {Puccetti}, {Ramsey}, {Rankin}, {Ratheesh}, {Roberts}, {Romani}, {Sgr{\`o}},
  {Slane}, {Soffitta}, {Spandre}, {Swartz}, {Tamagawa}, {Tavecchio}, {Tawara},
  {Tennant}, {Thomas}, {Tombesi}, {Trois}, {Turolla}, {Vink}, {Weisskopf},
  {Wu}, {Xie}, \& {Zane}}]{Tsygankov2023}
{Tsygankov}, S.~S., {Doroshenko}, V., {Mushtukov}, A.~A., {et~al.} 2023,
  \href{http://dx.doi.org/10.1051/0004-6361/202346134}{\JournalTitle{\aap},
  A48}

\bibitem[{{Ursini} {et~al.}(2023){Ursini}, {Marinucci}, {Matt}, {Bianchi},
  {Marin}, {Marshall}, {Middei}, {Poutanen}, {Rogantini}, {De Rosa}, {Di Gesu},
  {Garc{\'\i}a}, {Ingram}, {Kim}, {Krawczynski}, {Puccetti}, {Soffitta},
  {Svoboda}, {Tombesi}, {Weisskopf}, {Barnouin}, {Perri}, {Podgorny},
  {Ratheesh}, {Zaino}, {Agudo}, {Antonelli}, {Bachetti}, {Baldini},
  {Baumgartner}, {Bellazzini}, {Bongiorno}, {Bonino}, {Brez}, {Bucciantini},
  {Capitanio}, {Castellano}, {Cavazzuti}, {Ciprini}, {Costa}, {Del Monte}, {Di
  Lalla}, {Di Marco}, {Donnarumma}, {Doroshenko}, {Dovciak}, {Ehlert}, {Enoto},
  {Evangelista}, {Fabiani}, {Ferrazzoli}, {Gunji}, {Heyl}, {Iwakiri},
  {Jorstad}, {Karas}, {Kitaguchi}, {Kolodziejczak}, {La Monaca}, {Latronico},
  {Liodakis}, {Maldera}, {Manfreda}, {Marscher}, {Mitsuishi}, {Mizuno},
  {Muleri}, {Ng}, {O'Dell}, {Omodei}, {Oppedisano}, {Papitto}, {Pavlov},
  {Peirson}, {Pesce-Rollins}, {Petrucci}, {Pilia}, {Possenti}, {Ramsey},
  {Rankin}, {Romani}, {Sgr{\`o}}, {Slane}, {Spandre}, {Tamagawa}, {Tavecchio},
  {Taverna}, {Tawara}, {Tennant}, {Thomas}, {Trois}, {Tsygankov}, {Turolla},
  {Vink}, {Wu}, {Xie}, \& {Zane}}]{Ursini23}
{Ursini}, F., {Marinucci}, A., {Matt}, G., {et~al.} 2023,
  \href{http://dx.doi.org/10.1093/mnras/stac3189}{\JournalTitle{\mnras}, 519,
  50}

\bibitem[{{Veledina} {et~al.}(2024){Veledina}, {Muleri}, {Poutanen},
  {Podgorn{\'y}}, {Dov{\v{c}}iak}, {Capitanio}, {Churazov}, {De Rosa}, {Di
  Marco}, {Forsblom}, {Kaaret}, {Krawczynski}, {La Monaca}, {Loktev},
  {Lutovinov}, {Molkov}, {Mushtukov}, {Ratheesh}, {Rodriguez Cavero},
  {Steiner}, {Sunyaev}, {Tsygankov}, {Zdziarski}, {Bianchi}, {Bright},
  {Bursov}, {Costa}, {Egron}, {Garcia}, {Green}, {Gurwell}, {Ingram}, {Kajava},
  {Kale}, {Kraus}, {Malyshev}, {Marin}, {Matt}, {McCollough}, {Mereminskiy},
  {Nizhelsky}, {Piano}, {Pilia}, {Pittori}, {Rao}, {Righini}, {Soffitta},
  {Shevchenko}, {Svoboda}, {Tombesi}, {Trushkin}, {Tsybulev}, {Ursini},
  {Weisskopf}, {Wu}, {Agudo}, {Antonelli}, {Bachetti}, {Baldini},
  {Baumgartner}, {Bellazzini}, {Bongiorno}, {Bonino}, {Brez}, {Bucciantini},
  {Castellano}, {Cavazzuti}, {Chen}, {Ciprini}, {Del Monte}, {Di Gesu}, {Di
  Lalla}, {Donnarumma}, {Doroshenko}, {Ehlert}, {Enoto}, {Evangelista},
  {Fabiani}, {Ferrazzoli}, {Gunji}, {Hayashida}, {Heyl}, {Iwakiri}, {Jorstad},
  {Karas}, {Kislat}, {Kitaguchi}, {Kolodziejczak}, {Latronico}, {Liodakis},
  {Maldera}, {Manfreda}, {Marinucci}, {Marscher}, {Marshall}, {Massaro},
  {Mitsuishi}, {Mizuno}, {Negro}, {Ng}, {O'Dell}, {Omodei}, {Oppedisano},
  {Papitto}, {Pavlov}, {Peirson}, {Perri}, {Pesce-Rollins}, {Petrucci},
  {Possenti}, {Puccetti}, {Ramsey}, {Rankin}, {Roberts}, {Romani}, {Sgr{\`o}},
  {Slane}, {Spandre}, {Swartz}, {Tamagawa}, {Tavecchio}, {Taverna}, {Tawara},
  {Tennant}, {Thomas}, {Trois}, {Turolla}, {Vink}, {Xie}, \&
  {Zane}}]{Veledina24}
{Veledina}, A., {Muleri}, F., {Poutanen}, J., {et~al.} 2024,
  \href{http://dx.doi.org/10.1038/s41550-024-02294-9}{\JournalTitle{Nature
  Astronomy}, 8, 1031}

\bibitem[{{Weisskopf} {et~al.}(2022){Weisskopf}, {Soffitta}, {Baldini},
  {Ramsey}, {O'Dell}, {Romani}, {Matt}, {Deininger}, {Baumgartner},
  {Bellazzini}, {Costa}, {Kolodziejczak}, {Latronico}, {Marshall}, {Muleri},
  {Bongiorno}, {Tennant}, {Bucciantini}, {Dovciak}, {Marin}, {Marscher},
  {Poutanen}, {Slane}, {Turolla}, {Kalinowski}, {Di Marco}, {Fabiani},
  {Minuti}, {La Monaca}, {Pinchera}, {Rankin}, {Sgro'}, {Trois}, {Xie},
  {Alexander}, {Allen}, {Amici}, {Andersen}, {Antonelli}, {Antoniak},
  {Attina'}, {Barbanera}, {Bachetti}, {Baggett}, {Bladt}, {Brez}, {Bonino},
  {Boree}, {Borotto}, {Breeding}, {Brienza}, {Bygott}, {Caporale}, {Cardelli},
  {Carpentiero}, {Castellano}, {Castronuovo}, {Cavalli}, {Cavazzuti},
  {Ceccanti}, {Centrone}, {Citraro}, {D' Amico}, {D'Alba}, {Di Gesu}, {Del
  Monte}, {Dietz}, {Di Lalla}, {Di Persio}, {Dolan}, {Donnarumma},
  {Evangelista}, {Ferrant}, {Ferrazzoli}, {Ferrie}, {Footdale}, {Forsyth},
  {Foster}, {Garelick}, {Gunji}, {Gurnee}, {Head}, {Hibbard}, {Johnson},
  {Kelly}, {Kilaru}, {Lefevre}, {Le Roy}, {Loffredo}, {Lorenzi}, {Lucchesi},
  {Maddox}, {Magazzu}, {Maldera}, {Manfreda}, {Mangraviti}, {Marengo},
  {Marrocchesi}, {Massaro}, {Mauger}, {McCracken}, {McEachen}, {Mize}, {Mereu},
  {Mitchell}, {Mitsuishi}, {Morbidini}, {Mosti}, {Nasimi}, {Negri}, {Negro},
  {Nguyen}, {Nitschke}, {Nuti}, {Onizuka}, {Oppedisano}, {Orsini}, {Osborne},
  {Pacheco}, {Paggi}, {Painter}, {Pavelitz}, {Pentz}, {Piazzolla}, {Perri},
  {Pesce-Rollins}, {Peterson}, {Pilia}, {Profeti}, {Puccetti}, {Ranganathan},
  {Ratheesh}, {Reedy}, {Root}, {Rubini}, {Ruswick}, {Sanchez}, {Sarra},
  {Santoli}, {Scalise}, {Sciortino}, {Schroeder}, {Seek}, {Sosdian}, {Spandre},
  {Speegle}, {Tamagawa}, {Tardiola}, {Tobia}, {Thomas}, {Valerie}, {Vimercati},
  {Walden}, {Weddendorf}, {Wedmore}, {Welch}, {Zanetti}, \&
  {Zanetti}}]{Weisskopf2022}
{Weisskopf}, M.~C., {Soffitta}, P., {Baldini}, L., {et~al.} 2022,
  \href{http://dx.doi.org/10.1117/1.JATIS.8.2.026002}{\JournalTitle{JATIS}, 8,
  026002}

\bibitem[{{Wilms} {et~al.}(2000){Wilms}, {Allen}, \& {McCray}}]{Wilms2000}
{Wilms}, J., {Allen}, A., \& {McCray}, R. 2000,
  \href{http://dx.doi.org/10.1086/317016}{\JournalTitle{\apj}, 542, 914}

\bibitem[{{Wilson-Hodge} {et~al.}(2018){Wilson-Hodge}, {Malacaria}, {Jenke},
  {Jaisawal}, {Kerr}, {Wolff}, {Arzoumanian}, {Chakrabarty}, {Doty},
  {Gendreau}, {Guillot}, {Ho}, {LaMarr}, {Markwardt}, {{\"O}zel}, {Prigozhin},
  {Ray}, {Ramos-Lerate}, {Remillard}, {Strohmayer}, {Vezie}, {Wood}, \& {NICER
  Science Team}}]{WilsonHodge2018}
{Wilson-Hodge}, C.~A., {Malacaria}, C., {Jenke}, P.~A., {et~al.} 2018,
  \href{http://dx.doi.org/10.3847/1538-4357/aace60}{\JournalTitle{\apj}, 863,
  9}

\bibitem[{{Xiao} {et~al.}(2024){Xiao}, {Xu}, {Ge}, {Lu}, {Zhang}, {Zhang},
  {Tao}, {Qu}, {Wang}, {Kong}, {Tuo}, {You}, {Zhao}, {Peng}, {Du}, {Zhang}, \&
  {Ye}}]{Xiao24}
{Xiao}, Y.~X., {Xu}, Y.~J., {Ge}, M.~Y., {et~al.} 2024,
  \href{http://dx.doi.org/10.3847/1538-4357/ad24f8}{\JournalTitle{\apj}, 965,
  18}

\bibitem[{{Zel'dovich} \& {Shakura}(1969)}]{Zeldovich69}
{Zel'dovich}, Y.~B., \& {Shakura}, N.~I. 1969, 
  \href{https://ui.adsabs.harvard.edu/abs/1969SvA....13..175Z/abstract}{\JournalTitle{\sovast}, 13, 175}





\end{thebibliography}

\clearpage 
 
\appendix 
\onecolumn

\section{Optical polarimetric observations during 2017 outburst}
\label{sec:app1}

Table~\ref{tab:dipol} presents the results of the optical polarimetric measurements of J0243 during its 2017 outburst obtained with DIPol-2 \citep{Piirola2014} at the T60 telescope at Haleakala, Hawaii. 
Polarization of the field star \#3 (see Fig.~\ref{fig:sky}) was determined using observations in February 2024 at the same telescope. 

\begin{table*}[h]
\caption{Optical polarization of J0243 as observed with DIPol-2 in three filters \textit{B}, \textit{V}, and \textit{R} in 2017.} 
\centering
\begin{tabular}{lcccccc}
\hline
\hline
& \multicolumn{2}{c}{$B$} & \multicolumn{2}{c}{$V$} & \multicolumn{2}{c}{$R$} \\  
 HJD  & $q$ (\%) & $u$ (\%) &  $q$ (\%) & $u$ (\%) &  $q$ (\%) & $u$ (\%) \\
\hline
& \multicolumn{6}{c}{Observed polarization} \\ 
2458032.0946  & $-2.98\pm0.11$ & $-2.40\pm0.11$ & $-2.80\pm0.06$ & $-2.25\pm0.06$& $-2.92\pm0.06$ & $-2.32\pm0.06$ \\ 
2458033.0532  & $-3.23\pm0.08$ & $-2.22\pm0.08$ & $-3.02\pm0.08$ & $-2.18\pm0.08$& $-2.98\pm0.03$ & $-2.29\pm0.03$ \\ 
2458034.0270  & $-2.99\pm0.08$ & $-2.31\pm0.08$ & $-3.06\pm0.05$ & $-2.25\pm0.05$ & $-2.96\pm0.03$ & $-2.19\pm0.03$ \\   
2458037.0155  & $-3.07\pm0.07$ & $-2.13\pm0.07$ & $-2.97\pm0.05$ & $-2.31\pm0.05$ & $-2.92\pm0.03$ & $-2.27\pm0.03$ \\  
2458038.0295  & $-2.96\pm0.05$ & $-2.20\pm0.05$ & $-3.09\pm0.04$ & $-2.22\pm0.04$ & $-2.98\pm0.03$ & $-2.25\pm0.03$ \\  
2458043.9848  & $-2.91\pm0.08$ & $-2.03\pm0.08$ & $-2.88\pm0.12$ & $-2.28\pm0.12$ & $-2.87\pm0.04$ & $-2.17\pm0.04$ \\  
2458044.9984  & $-2.69\pm0.04$ & $-1.95\pm0.04$ & $-2.95\pm0.04$ & $-2.12\pm0.04$ & $-2.84\pm0.03$ & $-2.30\pm0.03$ \\    
2458045.9972  & $-2.94\pm0.05$ & $-2.32\pm0.05$ & $-3.02\pm0.05$ & $-2.18\pm0.05$ & $-2.88\pm0.03$ & $-2.20\pm0.03$ \\    
2458048.0195  & $-3.10\pm0.06$ & $-2.13\pm0.06$ & $-3.19\pm0.04$ & $-2.08\pm0.04$ & $-2.95\pm0.03$ & $-2.17\pm0.03$ \\    
2458053.9188  & $-3.03\pm0.05$ & $-2.46\pm0.05$ & $-3.08\pm0.04$ & $-2.22\pm0.04$ &  $-2.86\pm0.02$ & $-2.37\pm0.02$ \\    
2458054.9985  & $-2.93\pm0.04$ & $-2.41\pm0.04$ & $-3.09\pm0.03$ & $-2.17\pm0.03$ &  $-2.94\pm0.02$ & $-2.16\pm0.02$ \\    
2458056.0199  & $-3.02\pm0.03$ & $-2.26\pm0.03$ & $-3.03\pm0.03$ & $-2.20\pm0.03$ &  $-2.93\pm0.02$ & $-2.24\pm0.02$ \\    
2458057.0056  & $-2.97\pm0.04$ & $-2.30\pm0.04$ & $-3.02\pm0.04$ & $-2.20\pm0.04$ & $-2.99\pm0.03$ & $-2.24\pm0.03$ \\  
2458058.0394  & $-3.14\pm0.08$ & $-2.12\pm0.08$ & $-2.95\pm0.07$ & $-2.33\pm0.07$ & $-2.87\pm0.03$ & $-2.31\pm0.03$ \\  
2458060.9934  & $-3.01\pm0.05$ & $-2.18\pm0.05$ & $-2.94\pm0.04$ & $-2.26\pm0.04$ & $-2.89\pm0.02$ & $-2.29\pm0.02$ \\   
2458062.9920  & $-3.02\pm0.06$ & $-2.17\pm0.06$ & $-2.98\pm0.04$ & $-2.14\pm0.04$ & $-2.92\pm0.02$ & $-2.27\pm0.02$ \\  
2458063.9987  & $-3.09\pm0.05$ & $-2.28\pm0.05$ & $-3.03\pm0.04$ & $-2.22\pm0.04$ & $-2.97\pm0.02$ & $-2.29\pm0.02$ \\ 
2458064.9702  & $-2.94\pm0.03$ & $-2.41\pm0.03$ & $-3.05\pm0.03$ & $-2.24\pm0.03$ & $-2.95\pm0.02$ & $-2.34\pm0.02$ \\  
2458065.9628  & $-2.92\pm0.06$ & $-2.19\pm0.06$ & $-3.02\pm0.04$ & $-2.14\pm0.04$ & $-2.96\pm0.02$ & $-2.30\pm0.02$ \\  
2458067.9801  & $-2.73\pm0.03$ & $-2.28\pm0.03$ & $-2.95\pm0.04$ & $-2.30\pm0.04$ & $-2.84\pm0.02$ & $-2.34\pm0.02$ \\   
2458070.9535  & $-3.14\pm0.04$ & $-2.35\pm0.04$ & $-3.05\pm0.03$ & $-2.26\pm0.03$ & $-2.93\pm0.02$ & $-2.26\pm0.02$ \\ 
2458071.9512  & $-2.99\pm0.04$ & $-2.29\pm0.04$ & $-3.03\pm0.04$ & $-2.37\pm0.04$ & $-2.89\pm0.02$ & $-2.33\pm0.02$ \\  
2458074.9499  & $-3.06\pm0.05$ & $-2.23\pm0.06$ & $-3.02\pm0.04$ & $-2.21\pm0.04$ & $-2.91\pm0.02$ & $-2.35\pm0.02$ \\   
2458075.9985  & $-3.17\pm0.05$ & $-2.30\pm0.05$ & $-2.97\pm0.04$ & $-2.29\pm0.04$ & $-2.89\pm0.02$ & $-2.41\pm0.02$ \\    
2458077.8976  & $-3.05\pm0.05$ & $-2.10\pm0.05$ & $-2.96\pm0.06$ & $-2.39\pm0.06$ & $-2.82\pm0.03$ & $-2.32\pm0.03$ \\  
2458078.9242  & $-3.00\pm0.05$ & $-2.31\pm0.05$ & $-2.94\pm0.03$ & $-2.32\pm0.03$ & $-2.84\pm0.02$ & $-2.32\pm0.02$ \\ 
2458079.8771  & $-3.00\pm0.04$ & $-2.21\pm0.04$ & $-2.97\pm0.03$ & $-2.26\pm0.03$ & $-2.86\pm0.02$ & $-2.41\pm0.02$ \\  
2458081.9241  & $-3.01\pm0.06$ & $-2.25\pm0.06$ & $-3.01\pm0.05$ & $-2.22\pm0.05$ & $-2.88\pm0.02$ & $-2.36\pm0.02$ \\  
2458091.9003  & $-3.07\pm0.07$ & $-2.16\pm0.07$ & $-3.00\pm0.05$ & $-2.15\pm0.05$ & $-2.93\pm0.03$ & $-2.27\pm0.03$ \\   
2458093.8934  & $-3.04\pm0.07$ & $-2.03\pm0.07$ & $-3.08\pm0.05$ & $-2.20\pm0.05$  & $-2.87\pm0.03$ & $-2.31\pm0.03$ \\ 
2458094.9295  & $-3.11\pm0.06$ & $-2.13\pm0.06$ & $-3.06\pm0.06$ & $-2.23\pm0.06$  & $-2.90\pm0.02$ & $-2.31\pm0.02$ \\  
2458096.8955  & $-3.05\pm0.05$ & $-2.26\pm0.05$ & $-3.04\pm0.05$ & $-2.22\pm0.05$  & $-3.00\pm0.02$ & $-2.26\pm0.02$ \\  
2458097.8895  & $-3.08\pm0.05$ & $-2.22\pm0.05$ & $-3.04\pm0.04$ & $-2.21\pm0.04$  & $-3.02\pm0.02$ & $-2.28\pm0.02$ \\   
2458101.8661  & $-3.08\pm0.05$ & $-2.13\pm0.05$ & $-2.91\pm0.04$ & $-2.31\pm0.04$  & $-2.87\pm0.02$ & $-2.38\pm0.02$ \\    
2458112.8621  & $-2.85\pm0.04$ & $-2.21\pm0.04$ & $-3.03\pm0.05$ & $-2.27\pm0.05$  & $-2.91\pm0.02$ & $-2.29\pm0.02$ \\    
\hline
Average observed  & $-2.99\pm0.01$  & $-2.25\pm0.01$  & $-3.012\pm0.007$  & $-2.234\pm0.007$  & $-2.908\pm0.004$  & $-2.302\pm0.004$  \\ 
\hline
& \multicolumn{6}{c}{Interstellar polarization} \\  
Star \#3 & $-3.67\pm0.40$  & $-3.82\pm0.40$  &   $-2.79\pm0.30$ &   $-3.31\pm0.30$ &   $-2.81\pm0.20$  & $-3.11\pm0.20$   \\  
\hline 
& \multicolumn{6}{c}{Intrinsic polarization} \\ 
   & $\phantom{-}0.68\pm0.40$  & $\phantom{-}1.57\pm0.40$  &   $-0.22\pm0.30$ &   $\phantom{-}1.1\pm0.3$ &   $-0.1\pm0.2$  & $\phantom{-}0.8\pm0.2$ \\
\hline 
 & PD (\%) & PA (deg) & PD (\%) & PA (deg) &  PD (\%) & PA (deg) \\
 & $1.7\pm0.4$ & $33\pm7$& $1.1\pm0.3$ & $51\pm8$& $0.8\pm0.2$ & $49\pm7$ \\ 
\hline 
\end{tabular}
\tablefoot{Normalized Stokes parameters $q$ and $u$ are presented for the observed optical polarization of the source, the interstellar (IS) polarization, and the intrinsic polarization obtained by subtracting the IS polarization from the observed values. 
The PD and PA $\chi_{\rm o}$ of the intrinsic optical polarization are computed from the intrinsic $q$ and $u$.
Uncertainties are 1$\sigma$. }
\label{tab:dipol}
\end{table*} 

\end{document}